\documentclass[acmsmall]{acmart}

\usepackage{algorithm2e}
\usepackage{caption}
\usepackage{stfloats}
\usepackage{listings, listings-rust}
\usepackage{multirow}
\usepackage{xcolor}
\usepackage{colortbl}
\usepackage{array}
\newcolumntype{P}[1]{>{\centering\arraybackslash}p{#1}}

\usepackage{csquotes}
\MakeOuterQuote{"}

\AtBeginDocument{%
  }

\begin{document}

\title{NIFuzz: Estimating Quantified Information Flow with a Fuzzer}

\author{Daniel Blackwell}
\email{daniel.blackwell.14@ucl.ac.uk}
\orcid{0000-0001-7320-9057}
\affiliation{%
  \institution{University College London}
  \city{London}
  \country{UK}
}

\author{Ingolf Becker}
\email{i.becker@ucl.ac.uk}
\orcid{0000-0002-3963-4743}
\affiliation{%
  \institution{University College London}
  \city{London}
  \country{UK}
}

\author{David Clark}
\email{david.clark@ucl.ac.uk}
\orcid{0000-0002-7004-934X}
\affiliation{%
  \institution{University College London}
  \city{London}
  \country{UK}
}


\begin{abstract}
This paper presents a scalable, practical approach to quantifying information leaks in software; these errors are often overlooked and downplayed, but can seriously compromise security mechanisms such as address space layout randomisation (ASLR) and Pointer Authentication (PAC). We introduce approaches for three different metrics to estimate the size of information leaks, including a new derivation for the calculation of conditional mutual information.
Together, these metrics can inform of the relative safety of the target program against different threat models and provide useful details for finding the source of any leaks. We provide an implementation of a fuzzer, \qlfuzz, which is capable of dynamically computing these metrics with little overhead and has several strategies to optimise for the detection and quantification of information leaks.
We evaluate \qlfuzz{} on a set of 14 programs---including 8 real-world CVEs and ranging up to 278k lines of code in size---where we find that it is capable of detecting and providing good estimates for all of the known information leaks.
\end{abstract}

\def\qlfuzz{\textsc{NIFuzz}}
\newcommand{\cond}{\,|\,}

\maketitle

\section{Introduction}

\emph{Information flow control} (IFC) describes the controlled flow of information in systems. 
Typically, a failure of IFC that results in \emph{secret} information being revealed to users that lack sufficient privilege is referred to as an \emph{information leak}; such errors unfortunately go largely underreported.
One paper \cite{cho2020exploiting} --- which presents a technique for generating exploits from information leak vulnerabilities in OS kernels --- states that these ``vulnerabilities are usually assigned low CVSS [Common Vulnerability Scoring System] scores or [are left] without any CVEs [Common Vulnerability Enumerations] assigned''.

To give an example of where IFC is required, let's consider the context of a Linux-based OS where there is a file \verb|secrets.txt| for which only the user `root' has read privilege.
In this case, the OS is responsible for enforcing the IFC \emph{policy} that during OS execution information should not flow from \verb|secrets.txt| to any user other than `root'.

A shibboleth for IFC is some type of non-interference policy \cite{goguen1982security}.
For decades now this binary classification of flow policies (flow or non-flow) have been regarded as overly restrictive, i.e. small leaks may be tolerable depending on context, such as in programs like password checkers \cite{SZ04}.
When the leak is not tolerable, automated repair may require the size of the leak as part of a fitness function \cite{MesecanBCCP22}.
\emph{Quantified information flow} (QIF), as the name suggests, aims to quantify the amount of information flowing from one point to another, using information theory.
Typically it is used to quantify the amount of \emph{secret} information that is allowed to flow from privileged program input to unprivileged program output.
Taking the above example, if we were to execute the \verb|read| system call on \verb|secrets.txt| as some user other than `root' and received the output ``\texttt{P@ssword1}'', we would say that the quantity of information is 9 bytes \footnote{assuming each character is encoded by 1 byte}.
Here there is at least one unprivileged program input---the file descriptor for \verb|secrets.txt|---and also one privileged program input; that is the contents of the file itself. 
The privileged program input is provided by the kernel in response to the system call requesting the file contents, and this process is opaque to the unprivileged user.

Often, these \emph{information leaks} reveal only a fraction of the \emph{secret}; this is true for the OpenSSL Heartbleed bug which could leak up to $\approx64$KiB per request, regardless of the quantity of \emph{secret} data stored in the system.
The amount of information leaked can be dependent on both the privileged and unprivileged (\emph{secret}) inputs; for example, Heartbleed could leak between 0 and $\approx64$KiB depending on the unprivileged input -- which was the contents of the heartbeat request packet.
The maximum amount of information that can be leaked is referred to as the \emph{channel capacity} of the leak. 



In this paper, we present a new fuzzer \qlfuzz{} that can not only detect failures of IFC, but also provide a QIF estimate for them.
Where prior approaches to QIF estimation have scaling constraints \cite{heusser2010quantifying,phan2014abstract}, or do not provide automated program exploration \cite{chothia2013tool} --- both of which make automated analysis of large-scale software systems impossible --- our approach offers the scalability of fuzzing.
\qlfuzz{} provides both a strict lower-bound for channel capacity, as well as a state-of-the-art estimate of channel capacity for large leaks. 
Additionally we provide the first implementation of a testing-based approach that can measure of conditional mutual information between program inputs and outputs.
That is, the amount of leakage from \emph{secret} privileged inputs to unprivileged outputs, conditioned on the unprivileged inputs.
Conditional mutual information is a measure that is parameterised on the input distribution of the program, and gives the \emph{expected} amount of information leaked per execution; whereas the channel capacity gives the \emph{worst case} leakage.
\qlfuzz{} expands on concepts first introduced in LeakFuzzer \cite{blackwell2025hyperfuzzing}, and extends these to estimate QIF rather than merely detecting the presence of information leaks.

\qlfuzz{} is capable of determining whether the leakage has come from an \emph{explicit secret input}, as would be the case for the file contents example above, or \emph{stack} or \emph{heap} memory, as was the case for Heartbleed.
To the best of our knowledge, no fully-automated, scalable tool or approach currently exists that can monitor all three sources; let alone provide feedback on which source the \emph{secret} information came from.
The combination of the quantity of information leakage and its source can be extremely useful in locating and fixing IFC related bugs.
Larger leaks are likely to have more severe consequences, so awareness of the quantity of information leaked is useful for deciding on the priority of the bug.
Additionally QIF could be useful in confirming that programs which are intended to leak a certain amount of information, actually do leak that amount.
Finally, whilst our tool does not target side-channel leaks such as rowhammer \cite{kim2014flipping}, spectre \cite{kocher2019spectre} or meltdown \cite{lipp2018meltdown}; the QIF estimation approaches introduced in this paper could be applied to quantifying their channel capacities which contributes to their severity.

We evaluate \qlfuzz{} by generating QIF estimates for 14 programs containing information leaks, including 8 known CVEs, the largest program being 278k lines of code and containing leaks of up to 500,000+ physical bits.
On this corpus, \qlfuzz{} is able to identify all information leaks in all programs, and provide sensible lower-bounds and estimates for channel capacity in each case.








To summarise, the contributions of this paper are as follows:
\begin{itemize}
    \item Description of an approach for distinguishing the source of information leaks from explicit secret input, uninitialised stack- and uninitialised heap-memory.
    \item Description of an explore-exploit type mutation schedule to improve chances of discovering information leaks, and improve quantified information flow estimates.
    \item A publicly-available implementation of \qlfuzz{} including all of the above, and a evaluation of its leak detection and quantification estimating abilities.
\end{itemize}

\section{Background}

In this section, we provide a background overview of the two research areas from which \qlfuzz{} is drawn: fuzzing and information flow control.
Specifically, \qlfuzz{} is an application of fuzzing to the problem of information flow control.

\subsection{Fuzzing}


Fuzzing is a form of automated software testing that provides somewhat random inputs to a program and observes the outcome.
In general, the \emph{fuzzer} (the tool that orchestrates the fuzzing process) has limited observation power over the system under test (SUT) and commonly can only differentiate successful executions and crashes.
These crashes may be caused directly by bugs --- for example triggering a segmentation fault, or unhandled exception --- or assertion failures built in by the developer.

In the case of C and C++ programs, many bugs are caused by memory mismanagement such as \emph{use-after-free} or \emph{buffer overflows}.
These are typically undefined behaviour, but do not guarantee that the program will crash.
\emph{Sanitizers} are a group of compiler passes that add in runtime checks that force a crash when certain undefined behaviour occurs: \emph{AddressSanitizer} can detect out-of-bounds accesses, use-after-free, and similar; \emph{MemorySanitizer} can detect uninitialised memory reads; and \emph{ThreadSanitizer} can detect data races.

At first, fuzzers were simplistic and their main power came from the sheer rate of input generation and testing throughput; these \emph{black-box} fuzzers had no knowledge of program internals.
Black-box fuzzers, such as Zzuf \cite{hocevar_2007}, still have utility when fuzzing a target without access to the binary, such as a remote networked service.
Later came \emph{grey-box} and \emph{white-box} fuzzers; these have knowledge of the program structure and could leverage this to improve bug finding capabilities.
SAGE \cite{godefroid2012sage} is an example of a white-box fuzzer, it uses symbolic execution to allow for systematic testing of different execution paths; it does however have the program size constraints typical of symbolic execution.
AFL \cite{zalewski} is one of the best known grey-box fuzzers, and has a plethora of derivatives \cite{aflBasedFuzzers, fioraldi2020afl++, LibAFL, AFLFast}; it uses program coverage as feedback to guide the fuzzing process.
AFL generates new inputs to test by selecting from the corpus and applying mutations, Any generated input that discovers a previously uncovered control flow graph edge is appended to the corpus.

\subsection{Information Flow Control (IFC)}

In our context, we consider \emph{Information flow control} in terms of a software system; that is, we wish to ensure that the system adheres to some \emph{security policy}.
An early, and well known specification for security policies is the 1973 Bell-Lapadula model of \emph{access control} developed for the USA's military \cite{bell1976secure}.
The model has two mandatory \emph{access control rules}: a subject at a given level may not read an object at a higher security level, and a subject at a given level may not write to an object at a lower security level.
There is also a discretional access control rule, whereby access can be explicitly granted to specific users, allowing higher security level users to co-operate with a particular subset of lower security level users when the need arises.

\begin{figure}
  \includegraphics[width=0.7\columnwidth]{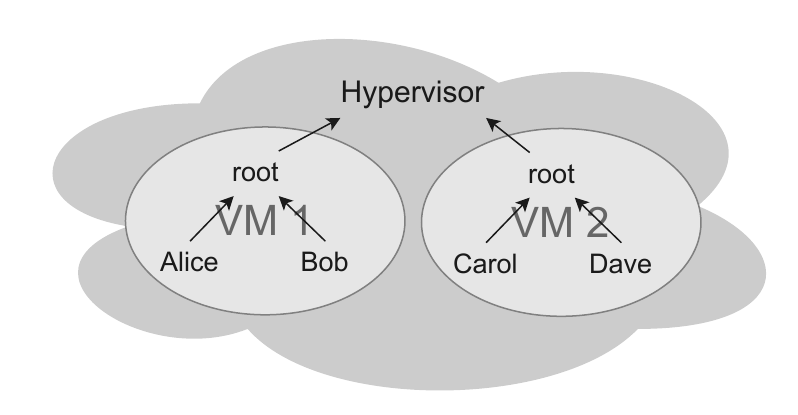}

  \caption{
    A simplified access control policy diagram for two isolated virtual machines running under a hypervisor; a common architecture for cloud computing services. 
    Here the arrow points represent allowed directions for flow of information.
    Note that both VM1 and VM2 have a `root' user, though these are separated and no information can flow between them.
  }
  \label{fig:lbac}
\end{figure}

This set of rules form a powerful system on their own, and later Denning realised that the ``no reads up, no writes down'' rule could be extended to a lattice ordering \cite{denning1976lattice}.
This \emph{lattice based access control} allows for more interesting policies with many users, an example of which is demonstrated in figure \ref{fig:lbac}.

The simplest useful lattice is the High-Low lattice, whereby information is allowed to flow from Low to High but not from High to Low.
We can reduce many of the flows in figure \ref{fig:lbac} to this, for example take Carol and Hypervisor; here Hypervisor maps to High and Carol maps to Low.
Information can flow from Carol to Hypervisor but not the other way.
This abstraction is used in \qlfuzz{}.

A \emph{security property} describes a property relating a program and \emph{security policy}.
A 1982 paper from Goguen and Meseguer \cite{goguen1982security} introduced the \emph{non-interference principle}:
\begin{quote}
A program P satisfies non-interference if and only if for any pair of low equivalent initial states, the resulting final states from running P with these initial states are also low equivalent.
\end{quote}

Note that the \emph{low} referred to here is synonymous with the Low in the High-Low lattice.
To visualise the \emph{non-interference principle}, let's take the following example function:

\begin{minipage}{\columnwidth}
  \begin{lstlisting}[language=C]
int sum(int low) {
  int high = getHighInput(); // read input from High user
  return low + high;
}
  \end{lstlisting}
\end{minipage}

Here the function will be treated as the complete program. 
We assume a security policy where information should not flow from \texttt{high} to \texttt{low}, and the function return value is visible to \texttt{low}. 
In order to demonstrate that the \emph{non-interference} property does not hold, we need a witness to its failure.
Given that this program is deterministic, a witness here needs only to observe different \emph{low} outputs for multiple executions with the same \emph{low} input.
Any difference in output must be due to the value of \emph{high} input.
Running the program with inputs \verb|{low: 0, high: 0}| gives output $0$, and running with \texttt{\{low: 0, high: 1\}} gives output $1$; hence the value of \texttt{high} input \emph{interferes} with \emph{low} output.
\qlfuzz{} identifies witnesses to the \emph{non-interference} property, and quantifies the amount of information that is revealed.

\subsubsection{Quantified Information Flow (QIF)}
\hfill \\
In this paper we measure the quantity of information flows with two measures; both of which are measures of Shannon information, and give a quantity in \emph{bits}.
The first of which measures the maximum amount of information that can be learnt from \emph{high} input from the \emph{low} output in a single execution, we refer to this as \emph{channel capacity} throughout our paper.
The second is the \emph{conditional mutual information}; that is, the mutual information between \emph{high} input and \emph{low} output, conditioned on the \emph{low} input.

In order to understand the \emph{channel capacity} measure, let's consider the following example, with four possible outputs $0, 1, 2, 3$ visible to \emph{low}: 

\begin{lstlisting}[language=C, morekeywords={uint8}]
uint8 target_func(uint8 low) {
  uint8 high = getHighInput(); // read input from High
  if (low % 4 == 0)
    return high % 4;
  else
    return low % 4;
}
\end{lstlisting}


We can calculate the \emph{channel capacity} described above by finding the value of \emph{low} input that allows for the most distinctions on \emph{low} output.
Given that this is a deterministic program, any differences in \emph{low} output must be caused by \emph{high} input (when the \emph{low} input is fixed).
A value of \emph{low} input for which there are most distinctions on output is $0$ (or any multiple of 4); this causes the return value to come from line 4, and this return value could be $0, 1, 2$ or 3.
We typically quantify the leakage in bits, which for the \emph{channel capacity} can be calculated by taking $log_2$ of the number of distinct low outputs; here that is $\log_2 4 = 2$ bits.

To calculate the \emph{conditional mutual information} measure, we can use the following formula \cite{coverthomas}:

\begin{equation}
    \label{eq:cmi}
    \begin{split}
    I(X;Y \cond Z) &= H(X \cond Z) - H(X \cond Y, Z) \\
    &= E_{P(x,y,z)} \log_2 \frac{P(X, Y \cond Z)}{P(X \cond Z)P(Y \cond Z)}
    \end{split}
\end{equation}

Where entropy of $X$, denoted $H(X)$, is defined as:

\begin{equation}
    H(X) = - \sum_{i=1}^{n} P(x_i) \log_2 P(x_i)
\end{equation}

And the conditional entropy of $X$ given $Z$, denoted $H(X \cond Z)$, is defined as:

\begin{equation}
    H(X | Z) = - \sum_{z \in Z} P(z) \sum_{x \in X} P(x | z) \log_2 P(x | z)
\end{equation}


Mapping \emph{low} output to $X$, \emph{high} input to $Y$ and \emph{low} input to $Z$ in Equation \ref{eq:cmi} gives us the mutual information between \emph{high} input and \emph{low} output conditioned on \emph{low} input, which works out as $0.5$ bits for the above listing and secure flow policy.
Without conditioning, the mutual information between \emph{high} input and \emph{low} output is only $\approx 0.0013$ bits.

There are many other ways to quantify information flow discussed in academic literature.
One uses the mutual information between \emph{high} input and \emph{low} output (without conditioning on the \emph{low} input) \cite{meng2011calculating}.
Another paper defines a \emph{conditional guessing entropy}, which describes the \emph{low} attacker's uncertainty of the \emph{high} input after making a guess and observing the result.

\subsubsection{Side-Channel Leakage}
\hfill \\
The leakage of secret information through channels other than by observing internal states or program output is termed \emph{side-channel leakage}.
These include observable metrics such as program execution time, memory consumption or power usage.
Cryptographic programs by their very nature must disclose all of the information contained within their secret inputs --- plaintext and encryption key --- and hence are not interesting for analysis for input-output leakage.
However, being able to learn information based on the side-channels allows an attacker to narrow the search space; thus much time and effort is put into ensuring that cryptographic implementations run in near-constant time.

\begin{figure}[H]
  \begin{lstlisting}[language=C, morekeywords={bool, String, min}]
bool checkPassword(String password, String guess){
  int iterations = min(password.len, guess.len);
  for (int i = 0; i < iterations; i++) {
    if (password[i] != guess[i])
      return false;
  }

  return password.length() == guess.length();
}
  \end{lstlisting}
\caption{A simple password checker containing a side-channel leak.}
\label{fig:sidechannel}
\end{figure}

A simple illustrative example is the password checker in figure~\ref{fig:sidechannel}.
We model the program with the actual \texttt{password} as a \emph{high} input, and the attacker's \texttt{guess} as \emph{low} input.
The number of passes around the loop in lines 3--6 depends on the number of correct characters in the guess.
As the attacker gets more and more characters correct, the runtime of the program increases.
If they were all letters, then worst case you need 26 guesses per character to get the correct one; so guessing a 9 letter password takes 9 * 26 = 234 guesses instead of $26^9$.
However, this is not the type of leakage that \qlfuzz{} works to detect, as it is not an input-output leak.
Instead, the return value does reveal some information about the secret \texttt{password} as each \texttt{false} (or indeed \texttt{true}) leaks information about the secret, but this is obviously a design necessity.

\section{Related Work}

The most closely related works are implementations of information flow control and quantified information flow; we have grouped these into four categories.
Firstly, static analysis approaches, then dynamic analysis approaches distinct from fuzzing.
Then fuzzing applied to side-channel leakage, and finally fuzzing applied to input-output leakage.

\subsection{Static Analysis}

An early technique capable of quantifying information flow in realistic programs is described in a 2007 paper by Clark, Hunt and Malacaria \cite{clark2007static}.
This paper uses a minimalist imperative language with static typing rules to illustrate the concepts, though no automated compiler was created and all of the example programs were manually checked.
A practical model-checking based approach to QIF was proposed in a 2010 paper by Heusser and Malacaria \cite{heusser2010quantifying}, this tool was evaluated 6 fragments of real-world programs containing information leak bugs and could provide lower-bounds on channel capacity up to 7 bits.
A 2014 paper by Phan and Malacaria \cite{phan2014abstract} provided an alternative model checking algorithm; this was able to produce the same bounds but in considerably less time.
As model checking techniques, the prior two works have scalability limitations which were overcame in a 2016 paper by Biondi et al. \cite{biondi2017hybrid}; this paper proposed a `hybrid' approach to QIF combining a precise program analysis with statistical methods in segments where precise analysis is too computationally complex.
It was evaluated on a set of 6 constructed programs, written in a simple programming language developed for the paper, ranging in size from 10 to 33 lines of code; and hence cannot be considered to be a `real world' approach in its describe form.

\subsection{Dynamic Analysis}


LeakiEst is a statistical tool for estimating QIF from a set of samples proposed in 2013; this operates on a simple text file containing pairs of (\emph{secret input}, \emph{low output}) and is thus language agnostic in contrast to the tools above.
The paper demonstrates that this is able to detect timing-channel leakage when provided the program runtime as output.
MutaFlow is a lightweight mutation-based analysis \cite{bjorn2017detecting} that mutates secret inputs at their source and observes whether a change has occurred at the program outputs.
It specifically targets android in the paper's evaluation, and achieves similar results to a static analysis tool FlowDroid \cite{arzt2014flowdroid} with much lower implementation cost (requiring only 10\% as many the source code lines).

\subsection{Fuzzing Applied to Side-Channel Leakage}


There are four papers introducing fuzzing-based tools to detect side-channel leakage, three of which target Java programs: JVM Fuzzing for JIT-Induced Side-Channel Detection \cite{brennan2020jvm}, DifFuzz \cite{nilizadeh2019diffuzz} and QFuzz \cite{noller2021qfuzz}.
DifFuzz does not attempt to quantify the leakage, only verify its presence, while the other two do provide estimates on leakage amount.
Both DifFuzz and QFuzz use a count of Java bytecode instructions to approximate program runtime, as opposed to using real execution time.
The fourth tool ct-fuzz \cite{he2020ctfuzz} also detects side-channel leaks, but in C and C++ programs; to do so it treats any path divergence due to secret input as leakage, and uses a model of CPU cache to approximate memory access delays; it counts the maximum number of possible paths for any public input, so can provide some notion of a lower-bound on the quantity of information leakage.

\subsection{Fuzzers}


LeakFuzzer is an application of fuzzing to detect input-output leaks in C and C++ programs \cite{blackwell2025hyperfuzzing}; it makes no attempt to quantify the leakage.
There is also a tool HyperFuzz (and a companion search-based software testing tool -- HyperEvo) described in a 2024 paper \cite{pasqua2024hypertesting} that can detect input-output leaks, but again not quantify them.


\section{Quantifying Information Flow}

Quantifying information is hard.
According to traditional information theory, more uncommon events contain more information.
In this work, we assume that all events follow a uniform distribution, as we do not know the real input distribution for the programs being tested.
Even with this assumption, quantifying information is still hard.
The reason that our approach can scale to arbitrarily large programs is that it is a testing approach -- meaning that only one input is tested at a time.
This unfortunately means that the only true way to measure information flow is to execute every possible input.
As this is not feasible for almost any interesting program, we must resort to estimating it instead.
There are three different estimation metrics that we update throughout the fuzzing campaign: channel capacity lower-bound, direct mappings and conditional mutual information.

\begin{figure}[H]
    \centering
    \includegraphics[width=\columnwidth]{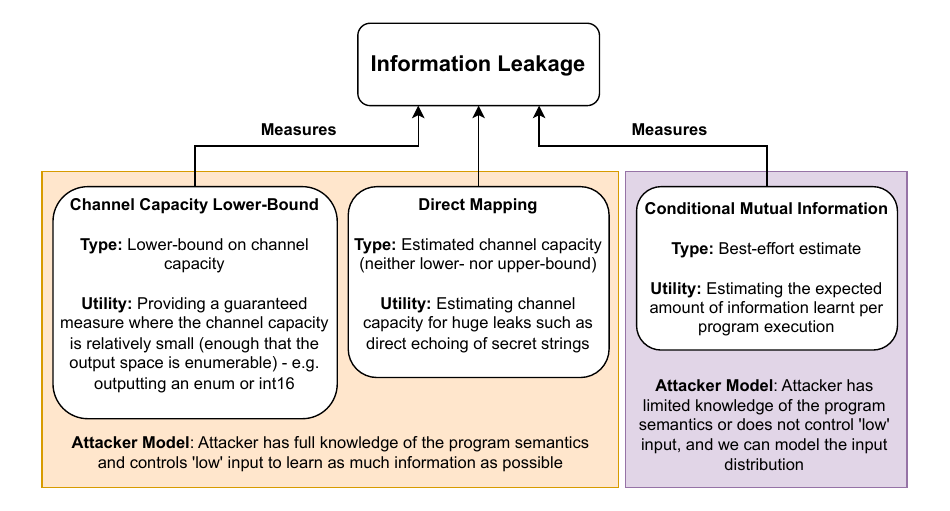}
    \caption{A diagram detailing the 3 approaches that we use for estimating quantifying information flow, note that the left two both measure channel capacity and all three are measures of Shannon information.}
    \label{fig:infoleaktypes}
\end{figure}



There are two attacker models that we consider.
The first of which is an attacker that has full knowledge of the program semantics and can control `low' input; the Heartbleed bug is such an example -- here the attacker controls the payload of the heartbeat packet which is sent to the server.
In such a case, we believe that the \emph{channel capacity lower-bound} and \emph{direct mapping} measures are the most useful, as it may reasonably be assumed that the attacker would choose a `low' input that maximises the amount of information learnt.

The second attacker model is of an attacker that either lacks knowledge of the program semantics or can observe but not control `low' input.
For this attacker model, we propose the estimated \emph{conditional mutual information} metric, which provides us with the amount of information learnt from `secret' input, conditioned on `low' input.

\subsection{Estimating Conditional Mutual Information}
\label{sec:calcingCMI}

The non-interference property guarantees that desired flow relations between the input partition and the output partition hold.
Not only can we use a non-interference property as an oracle to detect violations of such policies in software but we can use information theory to rank the severity of the violations, i.e. measure the degree of interference. 
Let $S^i$ be a sensitive input category in the data and $O^o$ be an output observation category of interest while $E^i$ (“Else”) is the union of all input parts apart from $S^i$, so $S^i \cup E^i = In$, the random variable in the inputs. 
A frequentist view of probability has it that Shannon's Conditional Mutual Information quantity expresses the information shared between $S^i$ and $O^o$, once one excludes any interference from the non-$S^i$ inputs, as 
\begin{equation}
  I(S^i; O^o \cond E^i).
\end{equation} 

In what follows we rewrite this quantity:

\begin{small}
  \begin{equation}
    \begin{array}{l@{}l}
I(S^i; O^o \cond E^i) &{}= H(S^i \cond E^i) - H(S^i \cond O^o, E^i) ~ \textrm{definition} \\
                   &{}= H(S^i, E^i) - H(E^i) - [H(S^i, O^o, E^i) - H(O^o, E^i)] ~ \textrm{chain rule} \\
                   &{}= H(In) - H(E^i) - [H(In, O^o) - H(O^o, E^i)] ~ \textrm{partition of In} \\
                   &{}= (H(O^o, E^i) - H(E^i)) - (H(In, O^o) - H(In)) \\
                   &{}= H(O^o \cond E^i) - H(O^o \cond In) ~ \textrm{chain rule}
    \end{array}
  \end{equation}
\end{small}

If the program is deterministic, the term $H(O^o \cond In) = 0$ as the output observations completely depend on the inputs so $I(S^i; O^o \cond E^i) = H(O^o \cond E^i)$.

When dealing with millions of inputs using a fuzzer, it is useful to minimise the amount of information that the fuzzer needs to record. 
Below we show that to calculate the interference in the deterministic case, a fuzzer need only record the violations of the non-interference property.
We have partitioned the inputs into $S^i \cup E^i = In$. 
Now we further partition the \emph{non-sensitive} input parts into those involved in violations and those not, $V \cup \overline{V} = E^i$.
Define $V$ as $V = \{e \in E^i \cond \exists s_1,s_2 \in S^i \wedge f(s_1,e) \neq f(s_2,e)\}$ where $f$ is the semantics of a deterministic program. 
$V$ induces a subset of the support for the random variable $\langle O^o,E^i \rangle$. 
Let us abuse notation slightly by using the name of the random variable as also the name for its support set.
Define $\langle O^o,E^i \rangle_V = \{(o,e)\in \langle O^o,E^i \rangle \cond e \in V\}$ as the subset of the support set for $\langle O^o,E^i \rangle$ that has a non-sensitive input appearing in one or more failing hypertests.
Assume that the program (semantics $f$) being fuzzed is deterministic in its behaviour.

\begin{equation}
  \begin{array}{l@{}l}
I(S^i; O^o \cond E^i) &{}= H(O^o \cond E^i) = H(O^o,E^i) - H(E^i) \\
                             &{}= - \sum_{(o,e)\in(O^o,E^i)} p(o, e) \log p(o, e) \\
                             &{} \indent + \sum_{e \in E^i} p(e) \log p(e) \\
                             &{}= - \sum_{(o,v)\in(O^o,E^i)_V} p(o, v) \log p(o, v) \\
                             &{} \indent - \sum_{(o,\overline{v}) \in (O^o,E^i)_{\overline{V}}} p(o, \overline{v}) \log p(o, \overline{v}) \\
                             &{} \indent + \sum_{v \in V} p(v) \log p(v) + \sum_{\overline{v} \in \overline{V}} p(\overline{v}) \log p(\overline{v}).
  \end{array}
\end{equation}
{}\\
Since the program is deterministic and $\bar{v}$ is not part of a violation, we have the following:
$\forall \bar{v} \in \overline{V} \,,\, \exists o \in O^o \,,\, \forall s \in S^i \,,\, f(s,\bar{v}) = o$. That is, if we pick a value for the non sensitive part of the input and that part is never involved in a violation, then as the secret part varies, the semantics will always produce the same output observation. So $p(\overline{v}) = p(o,\overline{v})$, since the probability distribution for $E^i$ is a marginal distribution of that for $\langle S_i,E^i \rangle$ and $o$ is fixed. Then we have
\begin{equation}
  -\sum_{(o,\overline{v}) \in (O^o,E^i)_{\overline{V}}} p(o, \overline{v}) \log p(o, \overline{v})
 + \sum_{\overline{v} \in \overline{V}} p(\overline{v}) \log p(\overline{v}) = 0
\end{equation} 
and
\begin{equation}
I(S^i; O^o \cond E^i) = - \sum_{(o,v) \in (O^o,E^i)_V} p(o,v) \log p(o,v) + \sum_{v \in V} p(v) \log p(v)
\end{equation}

Using this transformation, we can calculate the mutual information between secret input and low output, conditioned on low input, while keeping track \emph{only} of the hypertests that are violations of the security policy.
For programs where only a fraction of low inputs participate in a violation, this vastly reduces the amount of input-output mapping data that needs to be stored by the fuzzer and can make our estimated flow quantity more accurate.

\subsection{Implicit and Explicit Flows}

There are two types of input-output information flow; implicit and explicit.
Explicit flows are somewhat obvious, but implicit flows can be less intuitive; they are explained here to illustrate to ensure that the reader is aware of the different mechanisms by which information can flow.

In explicit flows, sensitive information is copied directly at some point, possibly to the program output by some form of assignment; for example:

\begin{lstlisting}[language=C]
int isEven(int input) {
  return input % 2;
}
\end{lstlisting}

Here we treat the return value as `output', the \verb|% 2| operation reduces the amount of information learnt to just 1 bit (compared to the 32 or 64-bit \verb|input|), but still explicitly returns a value directly derived from \verb|input|.
This explicit flow can pass through any number of variable assignments or operations:

\begin{lstlisting}[language=C]
int fifthBitNotSet(int input) {
  int shifted = input >> 5;
  int masked = shifted & 1;
  int isSet = masked;
  return !isSet;
}
\end{lstlisting}

This time we perform a more complicated set of operations, but information from \verb|input| still is explicitly passed through to the return value.

In implicit flows, information is learnt from the control flow of the program, for example:

\begin{minipage}{\linewidth}
\begin{lstlisting}[language=C]
const char *isEven(int input) {
  if (input % 2 == 1)
    return "false";
  else
    return "true";
}
\end{lstlisting}
\end{minipage}

In the above program, there is no direct \emph{explicit} assignment from \verb|input| to the return value; however receiving the return value \verb|"false"| \emph{implies} that the conditional check on line 2 evaluates to true, which in turn means that the value of \verb|input % 2| is 1.

All three of the above explicit and implicit flow examples leak exactly 1 bit of information, just through different means.

\subsection{Direct Mappings Between Input and Output}

\fbox{
    \begin{minipage}{0.95\columnwidth}
    \textbf{Note:} from this point onwards, we substitute the terms \emph{high} and \emph{low} that are typical in the theoretical IFC literature, with \emph{secret} and \emph{public}; where a typical security policy allows information to flow from \emph{public} to \emph{secret} but not vice versa.
    \end{minipage}
}
\vspace{2mm}

In order to calculate the channel capacity lower-bound by counting the number of unique public outputs for a given public input, 2\textsuperscript{\texttt{bits\_leaked}} outputs must be found.
For cases where the number of bits leaked is high, a true calculation of this quantity through this approach becomes infeasible -- to quantify 30 bits, $>1$ billion outputs must be found. However, in some programs, e.g. HeartBleed and other leaks from memory regions, the semantics of the vulnerability means that there is effectively a one to one mapping between the secret and the observations -- and we can exploit this directly.
If we can observe that flipping a bit in the input results in a bit in the output being flipped, we can approximate the size of these large leaks.

We propose building a mapping from \emph{secret} input bits to corresponding \emph{public} output bits.
We say that a mapping exists between a \emph{secret} input bit and a \emph{public} output bit if flipping the given input bit (either from 0 to 1 or vice versa), also results in the output bit being flipped.
The simplest trivial example of this is as follows, again we treat return value as output:

\begin{minipage}{\linewidth}
\begin{lstlisting}[language=C]
int identity(int input) {
  return input;
}
\end{lstlisting}
\end{minipage}

If we consider fixed precision \verb|int|s in their binary form, we can see that flipping any bit of \texttt{input} results in the same bit of the output being flipped. 
For example, the input \texttt{0b00} trivial produces the output \texttt{0b00}.
Flipping the first bit of the input gives us \texttt{0b01}, and the corresponding output \texttt{0b01}.
Instead, flipping the second bit of the input gives us \texttt{0b10} and the output \texttt{0b10}.
The mapping here can be extended to the full set of bits that make up \texttt{input}; if using 32-bit \texttt{int}s, this would require discovering $2^{32}$ ($\approx$4 billion) unique outputs.

As a trivial example, the above could be done by searching for exact matches between input and output bits.
To demonstrate the advantage of bitflips, let's consider a slightly trickier program:

\begin{minipage}{\linewidth}
\begin{lstlisting}[language=C]
int bitwiseNotted(int input) {
  return ~input;
}
\end{lstlisting}
\end{minipage}

Here we apply a bitwise NOT to the return value; the input \texttt{0b00} produces output \texttt{0b11...11} (all 32-bits are now set to 1) and input \texttt{0b01} produces output \texttt{0b11...10}.
If we were to simply search for occurrences of the provided input in the corresponding output, we would find none; but observing the position of flipped output bits still works.

This concept to allow fast approximations of quantities for large leaks is expanded upon in section \ref{bitflips} with the description of an algorithm to produce such a mapping automatically.

\section{Technique and its Implementation}

\qlfuzz{} has been implemented in Rust using components from LibAFL \cite{LibAFL}, and the source code is publicly available\footnote{\url{https://github.com/DanBlackwell/NIFuzz}}.
We have chosen to implement an input-output leakage measurement, however the technique could be applied to side-channels such as timing or power with an appropriate channel quantification methodology.
In our input-output implementation we use a forkserver to run the SUT (system under test) and inherit the \texttt{stdout} and \texttt{stderr}; this allows for a higher throughput than running the SUT as a command (via exec or similar).
Additionally we use the standard LibAFL coverage map feedback to guide the search, this stores any inputs that discover new branches to a corpus for further mutation.

The following sections discuss the specifics of \qlfuzz{} that diverge from a typical AFL-like fuzzer.
Firstly we discuss \emph{input structure} which is needed for the fuzzer to distinguish whether output changes are due to \emph{secret} input as opposed to \emph{public}.
Then we look at the \emph{mutation phase}, which is split into \emph{explore and exploit} subphases.
This is followed by the specifics of how \qlfuzz{} maximises the observable quantity of leakage using specific mutations.
Finally we discuss how QIF estimates are produced from the data stored by the fuzzer.


\section{Input Structure}

In order for the fuzzer to know that it has discovered a leak, it must have found two inputs with matching \emph{public} parts (and differing \emph{secret} parts) that produce differing outputs (assuming that the program is deterministic); therefore the fuzzer must be aware of which parts of the input are \emph{secret}.
Hence input structure is a key ingredient.

\subsection{Explicit Inputs, Stack Memory and Heap Memory Secrets}

\qlfuzz{} has the ability to determine whether leaked information has come from explicit \emph{secret} input, uninitialised stack memory or uninitialised heap memory.
This is beneficial as once a leak has been discovered, this information can be used to help pinpoint the program point at which the \emph{secret} information started being treated as \emph{public} information.

\begin{figure}
    \centering
    \includegraphics[width=0.8\linewidth]{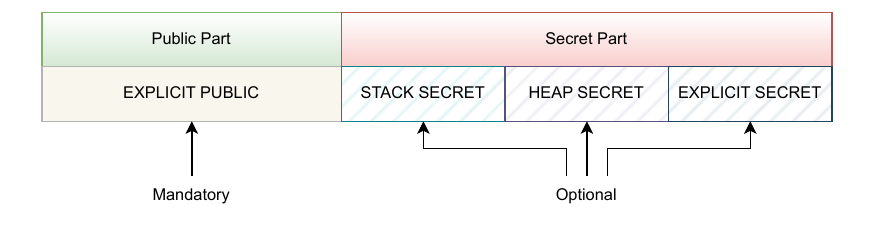}
    \caption{Diagram showing the structure of an input, as seen from \qlfuzz{} point of view. Note that each subsection of the secret part of the input is optional.}
    \label{fig:inputstructure}
\end{figure}

To make this possible, both the internal and external representation of program inputs have 1 \emph{public} and 3 distinct \emph{secret} input parts: explicit input, stack memory and heap memory.
We provide an easy method for retrieving each of these parts within the fuzzing harness, and function implementations to populate the uninitialised portions of stack\footnote{for x86 and AArch64} and heap memory with the provided values.

Additionally, it is possible for programs to only use the parts of the input that are necessary, for example some may not use explicit secret input.
\qlfuzz{} will not populate these parts of the input nor waste time attempting to mutate them.
Note that whilst it would be possible to run \qlfuzz{} with no secret parts at all, it would be better to use a standard grey-box fuzzer instead, as the mutation engine and corpus schedulers are optimised towards detecting information leaks.

To illustrate an example of stack memory leakage, the following code is taken from the Linux kernel (CVE-2009-2847) as used in our evaluation:

\begin{minipage}{\linewidth}
\begin{lstlisting}[language=C, morekeywords={size_t, stack_t, String, min}]
typedef struct sigaltstack {
  void __user *ss_sp;
  int ss_flags;
  size_t ss_size;
} stack_t;

int
do_sigaltstack (const stack_t __user *uss,
                stack_t __user *uoss, 
                unsigned long sp)
{
  stack_t oss;
  ...
  if (uoss) {
    oss.ss_sp = (void __user *) current->sas_ss_sp;
    oss.ss_size = current->sas_ss_size;
    oss.ss_flags = sas_ss_flags(sp);
  }
  ...
  if (uoss) {
    if (copy_to_user(uoss, &oss, sizeof(oss)))
  ...
  }
}
\end{lstlisting}
\end{minipage}

In the Linux kernel, variables marked with \verb|__user| (such as \verb|ss_sp| on line 2) come from untrusted sources; these are treated as \emph{public} input in our evaluation.
The function \verb|copy_to_user| on line 21 copies the memory contents of the kernel-space variable \texttt{oss} to the user-space \texttt{uoss}.
We can see that \verb|oss| is declared on line 12, but it is not initialised; however on lines 15--17 all three struct member values are populated.
A reasonable assumption would be that besides those three struct members, no further information is disclosed from kernel- to user-space.
In fact, there are 4-bytes of potentially sensitive kernel-space memory leaked due to struct \emph{padding}.
The way that the \verb|sigaltstack| (aka \verb|stack_t|) struct is layed out in memory is shown in the following diagram:

\begin{table}[H]
  \begin{footnotesize}
    \begin{tabular}{|l||P{0.8cm}P{0.8cm}||P{0.8cm}|P{0.8cm}||P{0.8cm}P{0.8cm}||}
    \hline
    \textbf{Mem. Addr.} & \multicolumn{1}{P{0.8cm}|}{0-3} & 4-7 & 8-11      & 12-15 & \multicolumn{1}{P{0.8cm}|}{16-19} & 20-23 \\ \hline
    \textbf{Variable}       & \multicolumn{2}{P{1.6cm}||}{ss\_sp}    & ss\_flags &       & \multicolumn{2}{P{1.6cm}||}{ss\_size}      \\ \hline
    \end{tabular}
  \end{footnotesize}
\end{table}

On 64-bit architectures, the compiler prefers to align variables to 8-byte (64-bit) boundaries, indicated here by double vertical lines, this is done for performance reasons.
Additionally, it must lay out struct members in the order that they are defined.
As \verb|ss_flags| is a 4-byte integer, the compiler inserts 4 bytes of padding after it, so as to align \verb|ss_size| with a boundary.
So despite setting values for all struct members, bytes 12-15 in the above diagram are never initialised, and therefore still hold the value of the variable previously stored in this area of memory.
Thus, when \verb|copy_to_user| is called, these 4-bytes --- containing potentially sensitive kernel-space information --- are copied out to user-space in the \verb|uoss| variable.
A simple fix in this case is to add a call to \verb|memset| the memory containing \verb|oss| to $0$ before populating the struct member values.

In this particular example, the memory containing the leaked information was on the stack, as \verb|oss| is a local function variable.
If instead \verb|oss| was allocated on the heap with \verb|malloc|, then it would have been heap memory contents leaked instead.
By detecting that the leak is 4-bytes, comes from the stack, and from a leak within this function, it should be possible to find the source of the bug and produce a patch.

To illustrate an example of explicit \emph{secret} input leakage, we use the following code taken from Fedora's NetworkManager (CVE-2011-1943) as used in our evaluation:

\begin{lstlisting}[language=C]
void
nm_setting_vpn_add_secret (NMSettingVPN *setting,
                           const char *key,
                           const char *secret)
{
  ...
  g_hash_table_insert (
    NM_SETTING_VPN_GET_PRIVATE (setting)->secrets,
    g_strdup (key), g_strdup (secret));
} 

static void
destroy_one_secret (gpointer data)
{
  char *secret = (char *) data;

  /* Don't leave the secret lying around in memory */
g_message ("%s: destroying %s", __func__, secret);
  memset (secret, 0, strlen (secret));
  g_free (secret);
}
\end{lstlisting}

Here the \verb|secret| parameter to the function \texttt{nm\_setting\_vpn\_add\_secret} is an explicit secret input; it is expected by the user that this secret should only be used within the NetworkManager system.
As seen on line 18, the secret is printed in plain-text by a logging function.
We would assume, due to the incorrect indentation and preceding comment, that this particular line of code was added for debugging and mistakenly made its way into release builds.

\section{Mutation Phases}

Now that we have covered the structure of the inputs, we can explain the mutation phase.
The mutation phase is split into an \emph{explore} stage that aims to detect \emph{violations} of the security policy (i.e. information leaks), and \emph{exploit} stage which aims to quantify these leaks.
The \emph{explore} stage is run exclusively until a violation is found, and then run interchangeably with the \emph{exploit} stage afterwards.

\subsection{Explore Stage (leak discovery)}

As hinted at above, the \emph{explore} stage is the default mode of \qlfuzz{}.
It aims to explore program functionality using the typical coverage-guided grey-box fuzzing methodology, but also search for the presence of a violation (that is, a \emph{public} input that produces different program \emph{public} output depending on the value of \emph{secret} input).
When we refer to \emph{public output}, we are referring to output which is visible to \emph{public} level users.
The following steps are repeated:

\begin{enumerate}
  \item Select an input from the fuzzer input corpus.
  \item Mutate the \emph{public} part of the input, execute the target program with this input and collect the output.
  \item Mutate the \emph{secret} parts of the input, execute the target program with this input and collect the output.
  \item Mutate the \emph{public} and \emph{secret} parts of the input, execute the target program with this input and collect the output.
\end{enumerate}
Both the \emph{public} and \emph{explicit secret} (if present) parts of the input independently have traditional fuzzing mutations applied, such as bitflips or splicing.
The \emph{stack memory secret} and \emph{heap memory secret} parts of the input have a simple mutation applied, in (3) they are set to \verb|0b10101010| and \verb|0b01010101|.
The functions used to fill uninitialised memory in the SUT repeat the relevant part of the input until the memory area is completely filled.
These two 1-byte inputs therefore fill the uninitialised memory with opposing bit values resulting in any direct disclosure from uninitialised memory producing a different observable output.

Note that we do not aggregate (2) and (3), as in a conventional fuzzer, since the program may be sensitive to secret inputs and also mutating public parts alone may be necessary to reach new coverage.
Imagine an input for the following program with \verb|{ secret: 1, public: 0 }|:

\begin{lstlisting}[language=C]
if (secret == 1) {
  if (public < 3) return 0;
  else return 1;
} else {
  return 0;
}
\end{lstlisting}


We could only reach the \emph{else} branch on line 3 by retaining the current value of \verb|secret| and mutating \verb|public| to be >= 3.

If a pair of tests with matching \emph{public} inputs, but differing \emph{public} outputs (due to \emph{secret} input rather than non-determinism), are found then these are stored and the \emph{public} input is added to the set of \emph{violations}.

We cache information about every input and output along the way in order to maximise our odds of detecting a violation.
The fuzzer state maintains a \verb|map| as shown in figure \ref{fig:statemap} (note that this object is simplified, and in our implementation, hashes are stored rather than the full byte array wherever possible). 

\begin{figure}[H]
  \begin{lstlisting}[language=Rust, morekeywords={PublicInput, IOHashValue, hashmap, OutputData, SecretInputParts, byte}]
  type OutputData {
    stdout: [byte],
    stderr: [byte]
  }

  type SecretInputParts {
    // the '?' here indicates an optional value, one or more members must be 
    // populated
    explicit_secret: [byte]?,
    stack_memory_secret: [byte]?,
    heap_memory_secret: [byte]?
  }
  
  // The IOHashValue object is defined as follows
  type IOHashValue {
    // each unique public output is stored with a corresponding secret input
    secret_input_for_public_output: HashMap<OutputData, SecretInputParts>,
  }

  // `PublicInput` is a type alias for list of `byte`
  typedef [byte] PublicInput;

  // This stores a mapping from 'public input' to an `IOHashValue` object
  map: hashmap<PublicInput, IOHashValue>
  \end{lstlisting}

  \caption{Simplified description of the \texttt{map} stored in the fuzzer state}
  \label{fig:statemap}
\end{figure}



\subsection{Exploit Stage (leak quantification)}

This stage operates on a violation, thus will not run until a violation has been discovered.
At a high-level, the stage operates as follows:

\begin{enumerate}
  \item Select an input from the set of violations.
  \item See whether we have checked for direct mappings from secret input bits to output bits (i.e. flipping a particular bit from the \emph{secret} input results in one or more output bits flipping).
  \item If no, then perform the set of checks to determine whether such mappings exist (see section \ref{bitflips}).
  \item If yes, then sample secret inputs from uniform and execute them in order to collect the output.
\end{enumerate}

\subsection{Bitflips}
\label{bitflips}

This is the first phase of mutations applied during the \emph{exploit} phase on a newly discovered violation.
The aim is to produce a direct map from secret input bits to public output bits; that is a mapping whereby flipping this bit in the input results in the one (or more) bit in the output being flipped.
Not all information leaks will have such a mapping. 
The search procedure is as follows:

For each secret part of the input present (from the possibilities \emph{explicit secret input}, \emph{stack memory secret} and \emph{heap memory secret}), flip all bits of this part and run the SUT with this input.
If the output is changed, then store all bits of this part to the list \emph{influences}. Note that each member of \emph{influences} is essentially a tuple made up of (secret\_input\_part, bit\_number).

Then, if the number of elements in \emph{influences} is $<$ 1000, we flip one bit at a time in order to find potential mappings from secret input to output bits (as described in algorithm~\ref{algo:slowBitflips}).
Otherwise there are more than 1000 potential bits to flip, so we use a faster approximation that needs $\lceil\log_2|influences|\rceil$ steps to estimate all mappings (algorithm~\ref{algo:fastBitflips}).
Figure~\ref{fig:quickBitflipExample} shows a worked example with a small program leaking bits 3 and 6 from secret input directly to output.


\RestyleAlgo{ruled}
\begin{small}
\begin{algorithm}
  \DontPrintSemicolon
  \KwIn{\texttt{SUT} - The program being tested}
  \KwIn{\texttt{input} - An array of bits making up the original program input}
  \KwIn{\texttt{influences} - The set of bit positions in the input that map directly to output bits}
  \KwOut{\texttt{bitflipMap} - A hashmap from secret input bit positions to public output bit positions}
  \vspace*{0.2cm}
  \texttt{originalOutput} $\gets$ \texttt{executeWithInput(SUT, input)}\;
  \texttt{seenBitflips} $\gets \phi$ \tcp*{Empty Set}
  \texttt{bitflipMap} $\gets \{\}$ \tcp*{Empty HashMap}
  \For{\texttt{i}~\textbf{\emph{in}}~\texttt{influences}}{
    \texttt{mutatedInput $\gets$ input}\;
    \texttt{mutatedInput[i] $\gets$ mutatedInput[i]~XOR~1}\;
    \texttt{output $\gets$ executeWithInput(SUT, mutatedInput)}\; 
    \texttt{outputFlips $\gets$ findFlippedBits(originalOutput, output)}\;
    \texttt{manyToOnes $\gets$ seenBitflips $\cap$ outputFlips}\;
    \For{\texttt{(\_, outputFlips)}~\textbf{\emph{in}}~\texttt{bitflipMap}}{
      \texttt{outputFlips $\gets$ outputFlips $\setminus$ manyToOnes}\;
    }
    \texttt{seenBitflips $\gets$ seenBitflips $\cup$ outputFlips}\;
    \texttt{bitflipMap[i] $\gets$ outputFlips $\setminus$ manyToOnes}\;
  }
  \texttt{bitflipMap $\gets$ bitflipMap.filter(lambda(k, v)\{|v| > 0\})}
  \vspace*{0.3cm}

  \caption{Simple algorithm for flipping 1 bit at a time in order to find potential mappings from secret input to output bits.}
  \label{algo:slowBitflips}
\end{algorithm}
\end{small}

\begin{small}
\begin{algorithm}[H]
\DontPrintSemicolon
\KwIn{\texttt{SUT} - The program being tested}
\KwIn{\texttt{input} - An array of bits making up the original program input}
\KwIn{\texttt{influences} - The set of bit positions in the input that map directly to output bits}
\KwOut{\texttt{bitflipMap} - A hashmap from secret input bit positions to public output bit positions}
\vspace*{0.2cm}
\begin{flushleft}
\texttt{originalOutput $\gets$ executeWithInput(SUT, input)}\\
\texttt{outputToInputMap $\gets$ \{\}} \tcp*{Empty HashMap}
\For{\texttt{i $\gets$ 0} \KwTo \texttt{originalOutput.len}}{
  \vspace*{0.07cm}
  \texttt{outputToInputMap[i] $\gets$ 0}\\
}
\For{\texttt{bit $\gets$ 1} \KwTo $\lceil\log_2(\texttt{influences.len})\rceil$}{
  \vspace*{0.07cm}
  \texttt{bitValue $\gets$ 1} << \texttt{(bit - 1)}\\
  \texttt{mutatedInput $\gets$ input}\\
  \tcc{Note that mutatedInput.length is the length in bits}
  \For{\texttt{index $\gets$ 0} \KwTo \texttt{mutatedInput.length - 1}}{
    \If{\texttt{(index / bitValue) \% 2 == 1}}{
      \texttt{mutatedInput[i] $\gets$ mutatedInput[i]~XOR~1}\\
    }
  }
  \texttt{output $\gets$ executeWithInput(SUT, mutatedInput)}\\
  \texttt{outputFlips $\gets$ findFlippedBits(originalOutput, output)}\\
  \For{\texttt{i~}\textbf{\emph{in}}\texttt{~outputFlips}}{
    \vspace*{0.07cm}
    \texttt{outputToInputMap[i] $\gets$ outputToInputMap[i] + bitValue}\;
  }
}
\texttt{bitflipMap $\gets$ \{:\}} \tcp*{Empty HashSet}
\For{\texttt{(outputBit, inputBit)}~\textbf{\emph{in}}~\texttt{outputToInputMap}}{
  \If{\texttt{inputBit > 0}}{
    \vspace*{0.07cm}
    \texttt{bitflipMap[inputBit] $\gets$ bitflipMap[inputBit] $\cup$ outputBit}\\
  }
}
\end{flushleft}

\caption{Fast algorithm for finding potential mappings from secret input bits to output bits}
\label{algo:fastBitflips}
\end{algorithm}
\end{small}


\begin{figure}[H]
  \rule[1ex]{\columnwidth}{0.8pt}
  \raggedright
  \textbf{Function Under Test}\hfill
  \begin{lstlisting}[language=C]
if (public == 0)
  return secret & 0b01001000
else
  return 0
  \end{lstlisting}

  \vspace{2mm}
  \textbf{Initial Values}\hfill
  \begin{lstlisting}[numbers=none]
input = { public: 0b0, secret: 0b00000000 }
originalOutput = 0b00000000
outputToInputMap = { : }
  \end{lstlisting}

  \vspace{2mm}
  \textbf{Values After Each \texttt{bit} Loop Iteration}\hfill
  \begin{table}[H]
    \begin{tabular}{l|lllll}
      \texttt{bit}               & \texttt{bitValue} & \texttt{mutatedInput} & \texttt{output}     & \texttt{outputFlips} & \texttt{outputToInputMap}                                         \\ \hline
      1                        & 1        & 0b10101010          & 0b00001000 & {[}3{]}     & \{ 3: 1 \} \\
      2                        & 2        & 0b11001100          & 0b01001000 & {[}3, 6{]}  & \{ 3: 3, 6: 2 \} \\
      3                        & 4        & 0b11110000          & 0b01000000 & {[}6{]}     & \{ 3: 3, 6: 6 \}
      \end{tabular}
  \end{table}

  \textbf{Final Result}\hfill
  \begin{flushleft}
    \small{
      after inverting the mapping: \verb|bitflipMap = { 3: { 3 }, 6: { 6 } }|. i.e. bits 3 and 6 of \verb|secret| map directly to output.
    }
  \end{flushleft}
  \rule[1ex]{\columnwidth}{0.8pt}
  
  \caption{A worked example demonstrating how Algorithm \ref{algo:fastBitflips} finds the mapping between \texttt{secret} input bits 3 and 6, and the output bits 3 and 6 in the Function Under Test.}
  \label{fig:quickBitflipExample}
\end{figure}

The issue with both algorithms for detecting mappings is that output bits could be affected by non-determinism, resulting in spurious mappings.
We observed this in Heartbleed, where a particular 16-byte block of memory would be populated with `random' values -- it is possible that these were generated using system time or \verb|/dev/urandom| or similar.
To overcome this we follow up with a series of tests with combinations of the mapped input bits flipped and check that we see the expected output.
Any bits that flip in the output that were not predicted by the mapping are filtered out of any other mappings, as there is a many-to-one relationship.
Any bits that were expected to flip in the output but did not are also filtered out.
These checks are performed with all mappings at once, then $\frac{3}{4}$ of all mappings, all the way down to $\frac{1}{8}$ of all mappings; these are selected from uniform and the sampling continues until the set of mappings is stable (i.e. none are being filtered).

\subsection{Extending Secret Inputs - after bitflips}

If there are any one-to-many mappings between secret input and public output bits, it is possible that the input is being copied repeatedly into the output up to some range.
This is particularly prevalent in the \emph{stack memory secret} and \emph{heap memory secret} where our initial mutations during the explore stage set the length to just 1 byte, which is repeatedly copied into memory by the fuzzing harness.
With a single byte input, there can be a maximum of one byte (8 bits) of leakage per run; extending the input allows us to quantify a greater amount of leakage.
In order to find how much to extend the input by, the mappings are iterated through to find the greatest difference between two output flip positions (for the same input bit).
Take for example the following program, and its corresponding map, generated using one of the prior algorithms:

\begin{minipage}{\columnwidth}
\begin{lstlisting}[language=C, morekeywords={size_t, stack_t, String, min, int32, byte, print}]
// int32 is a 32-bit (4 byte) integer type
int32 targetFunc(int32 public) {
  int32 x;
  if (public == 1) {
    return x;
  } else {
    return 0;
  }
}

int32 main(void) {
  int32 public = getPublicInputFromFuzzer();
  byte[] stackSecret = getStackSecretFromFuzzer();
  fillStack(stackSecret);
  print(targetFunc(public)); // output the result
  return 0;
}
\end{lstlisting}
\end{minipage}

After calling \texttt{fillStack} with (for example) \texttt{stackSecret = [0xAA]}, stack memory looks like: \texttt{\{0xAA, 0xAA, ..., 0xAA\}}.
The corresponding \verb|bitflipMap| is: \\
$\{ 0: \{0,8,16,24\}, 1: \{1,9,17,25\}, 2: \{2,10,18,26\}, 3: \{3,11,19,27\},\\ 4: \{4,12,20,28\}, 5: \{5,13,21,29\}, 6: \{6,14,22,30\}, 7: \{7,15,23,31\} \}$.
In order to find the length that we should extend the secret input \verb|stackSecret| by, we iterate through each value in the mapping (a set of mapped output bits), and subtract the smallest from the largest value.
For example for key $0$, the value is the set $\{0,8,16,24\}$, so the maximum difference is $24 - 0 = 24$.
We do this for each element in \verb|bitflipMap| and store the maximum difference.
In this example, every element has the same difference, 24.

After finding this maximum value, we then extend the relevant part of the secret input by this length by repeatedly copying the current buffer until it is filled.
For example, extending \verb|stackSecret = [0xAA]| by 24 bits it becomes \\\verb|[0xAA, 0xAA, 0xAA, 0xAA]|.

We then rerun the bitflip mapping algorithm; which in this case produces a one-to-one mapping $\{ 0: \{ 0 \}, 1: \{ 1 \}, \dots, 31: \{ 31 \} \}$.
Now, instead of having 8 bits mapped directly from \verb|stackSecret| to output, we have 32 bits; which is the ground truth channel capacity for the program.

\section{Uniform Sampling}

After the bitflips have been completed---whether any direct mappings have been discovered or not---the fuzzer moves onto the next input.
Once the violation corpus contains at least one input, the fuzzer has a random 50/50 chance of selecting the next input from either the main corpus (populated according to program coverage) or the violation corpus.

If a violation is selected for the second time, i.e. the bitflip map has already been generated, then we begin a new mutation process: uniform sampling.
This is important for the conditional mutual information calculation, as it uses probabilities, which can be more accurately estimated using sampling with a known distribution.
Additionally, we can use the number of distinct outputs observed during this process to provide a lower-bound on the \emph{channel capacity}.

Populating the secret parts of the input from uniform is done by applying a single mutator that we have implemented as part of \qlfuzz{}.
This mutator replaces each byte from the relevant parts of the input with values produced by a random generator.
By doing this, the length of each part of the input remains unaltered, only the values change.
We then execute the target program with this input and store the mapping from secret input to public output in a distinct hashmap for uniform sampled inputs (with each \emph{violation} having its own hashmap).

In our implementation, we run this uniform sampling mutation $2^{16}$ (65,536) times before allowing the fuzzer to select a new input from a corpus for fuzzing.



\begin{figure}
  \includegraphics[width=0.7\columnwidth]{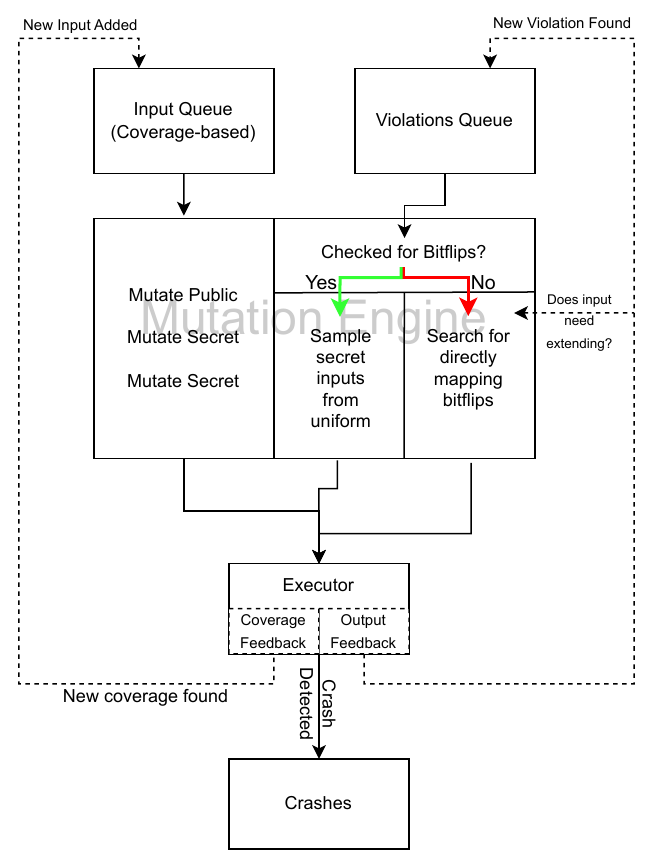}
  \caption{Lifecycle of an input in \qlfuzz{}.}
  \Description[
    A block diagram of the lifecycle of an input in \qlfuzz{}
  ]{
    A block diagram of the lifecycle of an input in \qlfuzz{}. Inputs travel from either the `input corpus' or `violations corpus' into the mutation engine, and the relevant mutation types are then applied. The mutated input is moved into the executor, coverage and output feedback collected and processed, and the mutated input may be added to the `input corpus' or `violation corpus' accordingly.
  }
\end{figure}

\section{Calculating QIF}

After finishing the mutation stage on a violation, we recalculate and output the quantified information flow statistics.
There are three values that we calculate:
\begin{enumerate}
  \item estimated conditional mutual information.
  \item channel capacity lower-bound; that is, the most information that can be learnt about the secret value from a single execution.
  \item Estimated most bits leaked directly to output; that is, the highest number of secret input bits that map directly to public output for a single violation.
\end{enumerate}

The way that these values are calculated is explained in the following subsections.

\subsection{Estimated Conditional Mutual Information}

For each public input, we store an \verb|IOHashValue| as was shown in figure \ref{fig:statemap}.
In order to estimate conditional mutual information, we store some extra fields on the object as is shown in Figure \ref{listing:IOHashValue}.

\begin{figure}
  \begin{lstlisting}[language=Rust, morekeywords={IOHashValue, OutputData, SecretInputParts}]
// dict stores a mapping from `public input' hash to an 
// IOHashValue object
map: HashMap<uint64, IOHashValue>

type IOHashValue {
  secret_input_for_public_output: 
        HashMap<OutputData, SecretInputParts>,
  // hits is incremented each time this public input 
  // gets run
  hits: uint64,
  // the following stores a map from public output hash 
  // to a list of secret input hashes that produced 
  // this output. But only for those generated by the 
  // uniform sampling mutator
  uniform_pub_outs_to_sec_ins: HashMap<uint64, [uint64]>,
  // The following stores the same, but for outputs 
  // generated during non-uniform stages of fuzzing (for 
  // example during the explore stage, or bitflips)
  non_uniform_pub_outs_to_sec_ins: HashMap<uint64, [uint64]>
}
  \end{lstlisting}

  \caption{Data structure for storing input-output mappings in fuzzer state}
  \label{listing:IOHashValue}
\end{figure}

Using these values we can calculate an estimate with the formula derived in section \ref{sec:calcingCMI}:
\begin{equation}
  I(S^i;O^o|E^i) = - \sum_{(o,v) \in (O^o, E^i)_V}p(o,v)\log p(o,v) + \sum_{v \in V} p(v) \log p(v)
\end{equation}

We will tackle the simpler second $\sum$ first; we can get all violations $V$ from the corpus specifically for storing \emph{violations}.
We can then calculate the probability $p(v)$ for each violation with $\frac{1}{number~of~unique~public~inputs}$.
The number of unique public inputs is the number of key-value pairs in \verb|map|.
We have found that more accurate estimates can be found earlier in the fuzzing campaign by only counting those public inputs that have more than a certain number of \verb|hits|.
This is likely to be because public inputs that have been sampled few times may be undetected \emph{violations}.

The first $\sum$ again requires only the violations corpus from the fuzzer, however we also need the set of public output--public input $(O^o, E^i)$ pairs.
To obtain these with the most accurate probabilities, we use the \verb|uniform_pub_outs_to_sec_ins| map for each of the violations.
We sum up the length of each list for each value in the map, giving us the total number of uniform samples for this violation.
We can then calculate $p(o,v)$ for each public output in the map (i.e. each key present), by multiplying the $p(v)$ value by the number of list elements and divide this by the total number of uniform samples.
As some public outputs may depend on very specific secret inputs --- and these may be near impossible to generate through uniform sampling, but have been found by the biased, heuristic-driven fuzzer mutation engine --- we also include any public outputs found in \verb|non_uniform_pub_outs_to_sec_ins|.
As we assume these to be rare, each is given a probability of 1 $\div$ `total number of uniform samples' for this violation.
While the number of uniform samples is low, these outputs will be over-represented (i.e. be treated as having a higher probability than they really do), however as more uniform samples are collected this error is reduced.

Using the above data structure, our approach is shown in algorithm~\ref{algo:cmiCalc}.

We additionally supply a command line flag that can force uniform selection of public inputs at all times; this is necessary for programs with large variations in branch probability.
Take, for example, a basic program having 2 paths with 1\% and 99\% probability respectively.
Due to the way that the fuzzer works by selecting at random from the input corpus, it is likely that the 1\% probability branch will be oversampled as whenever that input is selected from the corpus and mutated there is likely to be a much higher chance of generating a new input taking the same path.

\begin{small}
\begin{algorithm}
  \DontPrintSemicolon
  \texttt{uniquePublicInputs $\gets$ 0}\;
  \For{\texttt{(key,value)}~\textbf{\emph{in}}~\texttt{map}}{
    \tcc{Here, MIN\_HITS is some constant used to filter out undersampled violations}
    \If{\texttt{value.hits $>=$ MIN\_HITS}}{
      \texttt{uniquePublicInputs $\gets$ uniquePublicInputs + 1}\;
    }
  }

  \tcc{sumProbOutputs will contain $\sum_{(o,v) \in (O^o,E^i)_V} p(o,v) \log p(o,v)$}
  \texttt{sumProbOutputs $\gets$ 0}\;
  \tcc{sumProbViolations will contain $\sum_{v \in V} p(v) \log p(v)$}
  \texttt{sumProbViolations $\gets$ 0}\;
  \For{\texttt{v}~\textbf{\emph{in}}~\texttt{violations}}{
    \tcc{This is $p(v)$ (probability of sampling this violation)}
    \texttt{probViolation $\gets$ 1 $\div$ uniquePublicInputs}\;
    \texttt{sumProbViolations += probViolation $\cdot \log_2$(probViolation)}\;
    \texttt{hashVal $\gets$ map[v.publicInputHash()]}\;
    \texttt{uniformMap $\gets$ hashVal.uniform\_pub\_outs\_to\_sec\_ins}\;
    \texttt{numUniformSamples $\gets$ 0}\;
    \For{\texttt{(publicOutHash, secretInHashes)}~\textbf{\emph{in}}~\texttt{uniformMap}}{
      \texttt{numUniformSamples $\mathrel{+}=$ secretInHashes.len}\;
    }

    \For{\texttt{(publicOutHash, secretInHashes)}~\textbf{\emph{in}}~\texttt{uniformMap}}{
      \texttt{prob $\gets$ secretInHashes.len $\div$ numUniformSamples}\;
      \texttt{sumProbOutputs $\mathrel{+}=$ prob $\cdot \log_2$(prob)}\;
    }

    \texttt{uniformKeys $\gets$ uniformMap.keys()}\;
    \texttt{nonUniformKeys $\gets$ hashVal.non\_uniform\_pub\_outs\_to\_sec\_ins.keys()}\;
    \texttt{uniqueNonUniforms $\gets$ nonUniformKeys $\setminus$ uniformKeys}\;
    \For{\texttt{publicOutHash}~\textbf{\emph{in}}~\texttt{uniqueNonUniforms}}{
      \tcc{Assume that if we have not seen this output yet when sampling from uniform, then it's probability is 1 in numUniformSamples or less. To start with this overestimates the likelihood, but as numUniformSamples increases it becomes more accurate.}
      \texttt{prob $\gets$ 1 $\div$ numUniformSamples}\;
      \texttt{sumProbOutputs $\mathrel{+}=$ prob $\cdot \log_2$(prob)}\;
    }
  }

  \texttt{conditionalMutualInformation $\gets$ - sumProbOutputs + sumProbViolations}\;

  \caption{Our approach for estimating conditional mutual information using elements of the fuzzer state shown in figure~\ref{listing:IOHashValue}. This is an implementation of the equation $I(S^i; O^o \cond E^i) = - \sum_{(o,v) \in (O^o,E^i)_V} p(o,v) \log p(o,v) + \sum_{v \in V} p(v) \log p(v)$.}
  \label{algo:cmiCalc}
\end{algorithm}
\end{small}

\subsection{Channel Capacity Lower-Bound}

Channel capacity is the largest quantity of information that can leak from secret input to public output in a single execution of the target program.
This quantity of leakage is likely to only occur when certain inputs are provided. 

\qlfuzz{} produces a lower-bound for this particular value, as finding the exact quantity through search would require testing every public--secret input pair.
Again we use the fuzzer violation corpus, this time fetching the corresponding \verb|IOHashValue| for each violation.
We then find the number of elements in the set comprised of the union of the keys in the uniform and non-uniform maps from public output hash to secret input hash.
Finding the highest number of elements in any of these sets gives us the maximum number of distinctions on output (that we have discovered so far), and the $\log_2$ of this value is our lower-bound on channel capacity in bits.
A pseudocode description of this is shown in algorithm \ref{algo:maximalLeakage}.

\begin{small}
\begin{algorithm}
  \DontPrintSemicolon
  \texttt{maxOutputDistinctions $\gets$ 0}\;
  \For{\texttt{v}~\textbf{\emph{in}}~\texttt{violations}}{
    \texttt{hashVal $\gets$ map[v.publicInputHash()]}\;
    \texttt{uniformKeys $\gets$ hashVal.uniform\_pub\_outs\_to\_sec\_ins.keys()}\;
    \texttt{nonUniformKeys $\gets$ hashVal.non\_uniform\_pub\_outs\_to\_sec\_ins.keys()}\;
    \texttt{uniquePublicOutputs $\gets$ (uniformKeys $\cup$ nonUniformKeys).len}\;
    \If{\texttt{uniquePublicOutputs > maxOutputDistinctions}}{
      \texttt{maxOutputDistinctions $\gets$ uniquePublicOutputs}\;
    }
  }

  \caption{Our approach for producing a lower-bound on \emph{channel capacity} using the elements of fuzzer state shown in figure \ref{listing:IOHashValue}.}
  \label{algo:maximalLeakage}
\end{algorithm}
\end{small}

\subsection{Estimated Most Bits Leaked Directly to Output}

While our channel capacity lower-bounding approach can provide a guaranteed lower-bound, it requires at least as many executions as there are distinctions on public output.
For programs containing large leaks, such as the OpenSSL Heartbleed bug which leaks up to 64KiB per execution, it would take an infeasible amount of time and computer memory to attempt to produce a bound close to the true channel capacity.
The direct bitflip map between secret input and public output provides a way to estimate these large quantities.
Whilst we do a phase of testing combinations of these bitflips to search for interference between mappings, it would take the $2^{|\texttt{mappedBitflips}|}$ to exhaustively test them all; which would be the same as the number of distinctions on output if they did all map perfectly.
It is therefore possible that there may be interference between mappings that were not detected, and thus the true leakage is lower than the number of bitflips in our map.
It is also possible that there are more distinct outputs, for example:

\begin{lstlisting}[language=C]
  if (secret & 0b111 == 0b111)
    return 0b100
  else
    return secret & 0b11
\end{lstlisting}

In the above program, the input \verb|{secret: 0}| maps to the public output 0, \verb|{secret: 1}| maps to 1, and \verb|{secret: 2}| maps to 2.
In producing our map by flipping single bits, we would have the following map from secret input to public output: $\{ 0: \{0\}, 1: \{1\}\}$.
This would suggest a leak of 2 bits, or 4 distinctions on output; however there is a 5th when the input \verb|{secret: 7}| is given, where the output is 4.

Despite being an oversimplification, and capable of under- or over-approximation, the directly mapped bitflips can allow for a better estimation of leakage than the other estimators for very large leaks.

The bitflip map is stored in the \verb|IOHashValue|, so finding the estimated most bits leaked directly is simple; iterate through the violations corpus and find the number of elements in the largest map.



\section{Evaluation}

In this evaluation we aim to answer the following research questions:

\noindent \textbf{RQ1:} How does the lower-bound on channel capacity computed by \qlfuzz{} compare to the ground truth channel capacity for known information leaks?

\noindent \textbf{RQ2:} How does the estimate on channel capacity through direct input-output bit mapping compare to the ground truth channel capacity for known information leaks?

\noindent \textbf{RQ3:} How does the estimate on conditional mutual information perform on a program with known input distribution?

In order to answer these questions, \qlfuzz{} has been evaluated on a range of SUTs (Systems Under Test) including 8 information leak CVEs.
Channel capacity varies between 1 bit and in excess of 400,000 bits, and leaks come from explicit secret input, uninitialised stack memory and uninitialised heap memory.

As NIFuzz is a novel concept capable of scaling to program and leak sizes far beyond previous approaches for quantifying leaks in real-world C programs \cite{phan2014abstract,heusser2010quantifying}, but does not claim to have the guarantees of formal verification, it has been evaluated independently.

\begin{figure*}
 \begin{footnotesize}
  \begin{table}[H]
    \begin{tabular}{llll|lll}
    \multicolumn{1}{c}{\multirow{2}{*}{\textbf{SUT name}}} & \multicolumn{1}{c}{\multirow{2}{*}{\textbf{CVE number}}} & \multicolumn{1}{c}{\multirow{2}{*}{\textbf{Source}}} & \multicolumn{1}{c}{\multirow{2}{*}{\textbf{Lines of Code}}} & \multicolumn{3}{|c}{\textbf{Ground Truth Channel Capacity Bits}}                                                 \\
    \multicolumn{1}{c}{}                          & \multicolumn{1}{c}{}                            & \multicolumn{1}{c}{}                                 & \multicolumn{1}{c}{}                               & \multicolumn{1}{|c}{\emph{Explicit Input}} & \multicolumn{1}{c}{\emph{Stack Mem.}} & \multicolumn{1}{c}{\emph{Heap Mem.}} \\
    \rowcolor[HTML]{EFEFEF}  \texttt{AppleTalk}                                       & CVE-2009-3002                                    & \cite{heusser2010quantifying}                                    & 149                                                 & \multicolumn{1}{l}{-}              & \multicolumn{1}{l}{80}         & 0                              \\
    \texttt{cpuset}                                          & CVE-2007-2875                                    & \cite{heusser2010quantifying}                                    & 80                                                  & \multicolumn{1}{l}{-}              & \multicolumn{1}{l}{0}          & $2^{31}$                       \\
    \rowcolor[HTML]{EFEFEF}  \texttt{sigaltstack}                                     & CVE-2009-2847                                    & \cite{heusser2010quantifying}                                    & 110                                                 & \multicolumn{1}{l}{-}              & \multicolumn{1}{l}{32}         & 0                              \\
    \texttt{tcf\_fill\_node}                                 & CVE-2009-3612                                    & \cite{heusser2010quantifying}                                    & 788                                                 & \multicolumn{1}{l}{-}              & \multicolumn{1}{l}{16}         & 0                              \\
    \rowcolor[HTML]{EFEFEF}  \texttt{getsockopt}                                      & CVE-2011-1078                                    & \cite{phan2014abstract}                                       & 142                                                 & \multicolumn{1}{l}{-}              & \multicolumn{1}{l}{8}          & 0                              \\
    \texttt{Heartbleed}                                      & CVE-2014-0160                                    & \cite{fuzzertestsuite}                                     & 278,907                                             & \multicolumn{1}{l}{-}              & \multicolumn{1}{l}{0}          & \textgreater{}500,000          \\
    \rowcolor[HTML]{EFEFEF}  \texttt{NetworkManager}                                  & CVE-2011-1943                                    & \cite{blackwell2025hyperfuzzing}                                            & 115,643                                             & \multicolumn{1}{l}{320}            & \multicolumn{1}{l}{0}          & 0                              \\
    \texttt{RDS}                                             & CVE-2019-16714                                   & \cite{blackwell2025hyperfuzzing}                                            & 69,163                                              & \multicolumn{1}{l}{-}              & \multicolumn{1}{l}{16}         & 0                              \\
    \rowcolor[HTML]{EFEFEF}  \texttt{IFSpec\_Banking}                                 & -                                                & \cite{hamann2018uniform}                                          & 97                                                  & \multicolumn{1}{l}{1}              & \multicolumn{1}{l}{0}          & 0                              \\
    \texttt{IFSpec\_Password}                                & -                                                & \cite{hamann2018uniform}                                          & 66                                                  & \multicolumn{1}{l}{16}             & \multicolumn{1}{l}{0}          & 0                              \\
    \rowcolor[HTML]{EFEFEF}  \texttt{IFSpec\_Reviewers}                               & -                                                & \cite{hamann2018uniform}                                          & 105                                                 & \multicolumn{1}{l}{$\sim$18.5}       & \multicolumn{1}{l}{0}          & 0                              \\
    \texttt{explicit\_secret\_701\_bit}                      & -                                                & New                                                   & 34                                                  & \multicolumn{1}{l}{701}            & \multicolumn{1}{l}{0}          & 0                              \\
    \rowcolor[HTML]{EFEFEF}  \texttt{heap\_4808\_bit}                                 & -                                                & New                                                   & 34                                                  & \multicolumn{1}{l}{-}              & \multicolumn{1}{l}{0}          & 4808                           \\
    \texttt{stack\_17768\_bit}                               & -                                                & New                                                   & 33                                                  & \multicolumn{1}{l}{-}              & \multicolumn{1}{l}{17768}      & 0                              \\
    \end{tabular}
  \end{table}
 \end{footnotesize}

  \caption{
    SUT name is chosen arbitrarily based on the functionality of the tested code. 
    Lines of code was calculated using \texttt{cloc}, and counts the number of `C', `C++' and `C/C++ header' code lines. 
    IFSpec\_* benchmarks were translated manually from Java to C.
    All leakage is for code compiled with gcc targeting x86\_64 CPU, otherwise leakage from struct padding bytes may vary.
  }
\end{figure*}

\subsection*{Testing Environment}

All tests were run inside Docker containers based on the Ubuntu 20.04 distribution of Linux. 
Fuzzing experiments were run with 10 fuzzing campaigns in parallel, each bound to a single CPU core.
Tests were run on a dedicated server equipped with 2 x Intel Xeon E5-2620 v2 processors, making for a total of 12 cores (24 threads) at 2.10GHz, and 128GB RAM. 
Each benchmark was fuzzed for 12 hours.
Each experimental setup/run was repeated 10 times. 
The nondeterministic nature of the mutation engine results can differ greatly over 12 hours.

\subsection{Experimental Results}

A summary of results is shown in figure \ref{fig:resultsMain}.

\vspace{2mm}\noindent \textbf{RQ1:} How does the lower-bound on channel capacity computed by \qlfuzz{} compare to the ground truth channel capacity for known information leaks?

\vspace{2mm}\noindent For the programs with relatively small leaks (< 20 bits) -- \texttt{tcf\_fill\_node} (16 bits), \texttt{getsockopt} (8 bits), \texttt{RDS} (16 bits), \texttt{IFSpec\_Banking} (1 bit), \texttt{IFSpec\_Password} (16 bits) and \texttt{IFSpec\_Reviewers} ($\approx 18.5$ bits) -- the median lower-bound found in the runs is generally close to the ground truth.
Due to the relatively few number of outputs needing enumeration, the ground truth was discovered for \texttt{getsockopt} and \texttt{IFSpec\_Banking} in all runs.
In \texttt{RDS}, all runs came extremely close, finding at least 15.96 out of the 16 bits with the closest finding 15.9999 bits.

The largest discrepancy amongst the small leaks was \texttt{IFSpec\_Password}, which is due to the fuzzing harness.
The \emph{public} input in this program is a series of password guesses; in order to separate the single string provided by the fuzzer into substrings, we split on \verb|\n| characters.
The output observed is proportional to the number of guesses, with the program outputting ``No more password tries allowed'' for each guess after the \emph{secret} value \texttt{maxTries} has been reached.
As a result, in order to reach the true channel capacity, a \emph{public} input containing >65,536 \verb|\n| characters would need to be generated; which is very unlikely unfortunately.

The highest amount measured in any run is 19.5 bits in one run on \texttt{NetworkManager}; in that run there was a single violation for which 738,645 distinct outputs were found (by mutating the \emph{secret} part of the input).
This is still far below the ground truth lower-bound of 320 bits, and thus serves as good motivation for the coarser direct mapping approach for estimating channel capacity.

\vspace{2mm}\noindent \textbf{RQ2:} How does the estimate on channel capacity through direct input-output bit mapping compare to the ground truth channel capacity for known information leaks?

\vspace{2mm}\noindent We see that for 10 of the 14 programs, the estimate on channel capacity is much higher than the lower-bound discussed previously.
For one of the other 4 programs, \texttt{getsockopt}, both estimators produced the ground truth value.

For the three constructed programs---\texttt{expl\_sec\_701\_bit}, \texttt{heap\_4808\_bit} and \texttt{stack\_17768\_bit}---every run found the ground truth values for channel capacity.
These act as a sanity check to suggest that there is no over- or under-estimation from the algorithm in ideal circumstances, even when dealing with huge leaks.

For 5 of the 8 real-world CVEs, the median channel capacity discovered matches the ground truth exactly.
The largest of these is the 80 bit leak from uninitialised stack memory that occurs in \texttt{AppleTalk}.
For the other 3 programs, we still obtain estimates much higher than the provided lower-bounds and most notably a 443,960 bit estimate in the median run on \texttt{Heartbleed} (equating to 55,495 bytes against a 65,535 byte ground truth).

Finally, we see that for the \texttt{IFSpec\_\*} programs, this channel capacity estimate is somewhat ineffective.
This is due to the processing steps between input and output resulting in less direct correspondence between input and output bits.
Nonetheless, this does provide justification for computing both the lower-bound and direct mapping metrics simultaneously, as either can be more appropriate dependending on the leak.

\begin{figure}
\begin{table}[H]
  \begin{scriptsize}
  \begin{tabular}{l|p{0.4cm}p{0.4cm}p{0.4cm}|p{0.5cm}p{0.5cm}p{0.5cm}|p{0.4cm}p{0.6cm}p{0.7cm}|p{0.5cm}p{0.5cm}p{0.8cm}}
  \multicolumn{1}{c|}{\multirow{2}{*}{\textbf{SUT}}} & \multicolumn{9}{c}{\textbf{Direct Mapped Bitflips}}                                                                                 & \multicolumn{3}{|c}{\multirow{2}{*}{\textbf{Channel Capacity (Bits)}}} \\
  \multicolumn{1}{c|}{}                              & \multicolumn{3}{c}{\textit{Explicit Secret}} & \multicolumn{3}{c}{\textit{Stack Memory}} & \multicolumn{3}{c}{\textit{Heap Memory}} & \multicolumn{3}{|c}{}                                                 \\
  \rowcolor[HTML]{EFEFEF}  \texttt{AppleTalk}                                         & 0            & 0              & 0            & 80           & 80           & 80          & 0          & 0            & 0            & 16.3                 & 16.5                  & 17.1                  \\
  \texttt{cpuset}                                            & 0            & 0              & 0            & 0            & 0            & 0           & 8          & 128          & 128          & 8                    & 13.9                  & 18.8                  \\
  \rowcolor[HTML]{EFEFEF}  \texttt{sigaltstack}                                       & 0            & 0              & 0            & 32           & 32           & 32          & 0          & 0            & 0            & 16.6                 & 17.7                  & 18.9                  \\
  \texttt{tcf\_fill\_node}                                   & 0            & 0              & 0            & 0            & 0            & 0           & 16         & 16           & 16           & 11.9                 & 15.3                  & 15.7                  \\
  \rowcolor[HTML]{EFEFEF}  \texttt{getsockopt}                                        & 0            & 0              & 0            & 8            & 8            & 8           & 0          & 0            & 0            & 8                    & 8                     & 8                     \\
  \texttt{Heartbleed}                                        & 0            & 0              & 0            & 0            & 0            & 0           & 0          & 443,960      & 517,440      & 0                    & 16                    & 17.7                  \\
  \rowcolor[HTML]{EFEFEF}  \texttt{NetworkManager}                                    & 107          & 121.5          & 135          & 0            & 0            & 0           & 0          & 0            & 0            & 17.9                 & 18.5                  & 19.5                  \\
  \texttt{RDS}                                               & 0            & 0              & 0            & 16           & 16           & 16          & 0          & 0            & 0            & 15.96                & 15.999                & 15.9999               \\
  \rowcolor[HTML]{EFEFEF}  \texttt{IFSpec\_Banking}                                   & 0            & 0              & 1            & 0            & 0            & 0           & 0          & 0            & 0            & 1                    & 1                     & 1                     \\
  \texttt{IFSpec\_Password}                                  & 1            & 1              & 1            & 0            & 0            & 0           & 0          & 0            & 0            & 4.75                 & 5.02                  & 5.43                  \\
  \rowcolor[HTML]{EFEFEF}  \texttt{IFSpec\_Reviewers}                                 & 6            & 7.5            & 19           & 0            & 0            & 0           & 0          & 0            & 0            & 14.8                 & 16.1                  & 17.2                  \\
  \texttt{expl\_sec\_701\_bit}                        & 701          & 701            & 701          & 0            & 0            & 0           & 0          & 0            & 0            & 15.8                 & 16.8                  & 17.5                  \\
  \rowcolor[HTML]{EFEFEF}  \texttt{heap\_4808\_bit}                                   & 0            & 0              & 0            & 0            & 0            & 0           & 4,808      & 4,808        & 4,808         & 14.7                 & 15.2                  & 17.1                  \\
  \texttt{stack\_17768\_bit}                                 & 0            & 0              & 0            & 17,768       & 17,768       & 17,768      & 0          & 0            & 0            & 14.7                 & 15.1                  & 15.5                  
  \end{tabular}
  \end{scriptsize}
  \end{table}
  
\caption{
  Each cell gives results from the 10 runs in the format `min median max', with all values being rounded to 3 significant figures, except for NetworkManager channel capacity. 
  This is because all runs round to 16.0 bits (the ground truth), but none achieved exactly that figure.
  Note that \emph{Channel Capacity} here is the lower-bound that is calculated by taking log$_2$ of the number of distinct outputs for a single public input.
}
\label{fig:resultsMain}
\end{figure}

\vspace{2mm}\noindent \textbf{RQ3:} How does the estimate on conditional mutual information perform on a program with known input distribution?

\vspace{2mm}\noindent We provide a proof-of-concept level evaluation of CMI estimation, enabling the \qlfuzz{} flag to force uniform selection of public inputs. 
This allows us to make good estimates of programs with paths of vastly differing weight, such as the following which is used in our evaluation here:

\begin{minipage}{\linewidth}
\begin{lstlisting}[language=C, morekeywords={uint}]
typedef unsigned int uint32;
uint target_func(uint32 public, uint32 secret) {
  if (public % 10 == 0) {
    if (secret < 0xFFFF) return 0;
    else return 1;
  } else {
    return 101;
  }
}
\end{lstlisting}
\end{minipage}

Here, 1 in 10 public values will be part of the set of \emph{violations}; and for these $\frac{\texttt{0xFFFF}}{\texttt{UINT32\_MAX}}$ ($= \frac{1}{65537}$) values of secret will result in a return value of 1, while the rest return 0.
As a result, the mutual information between secret input and public output, conditioned on public input is just $0.0000266$ bits (3sf).

\begin{figure}
  \includegraphics[width=\columnwidth]{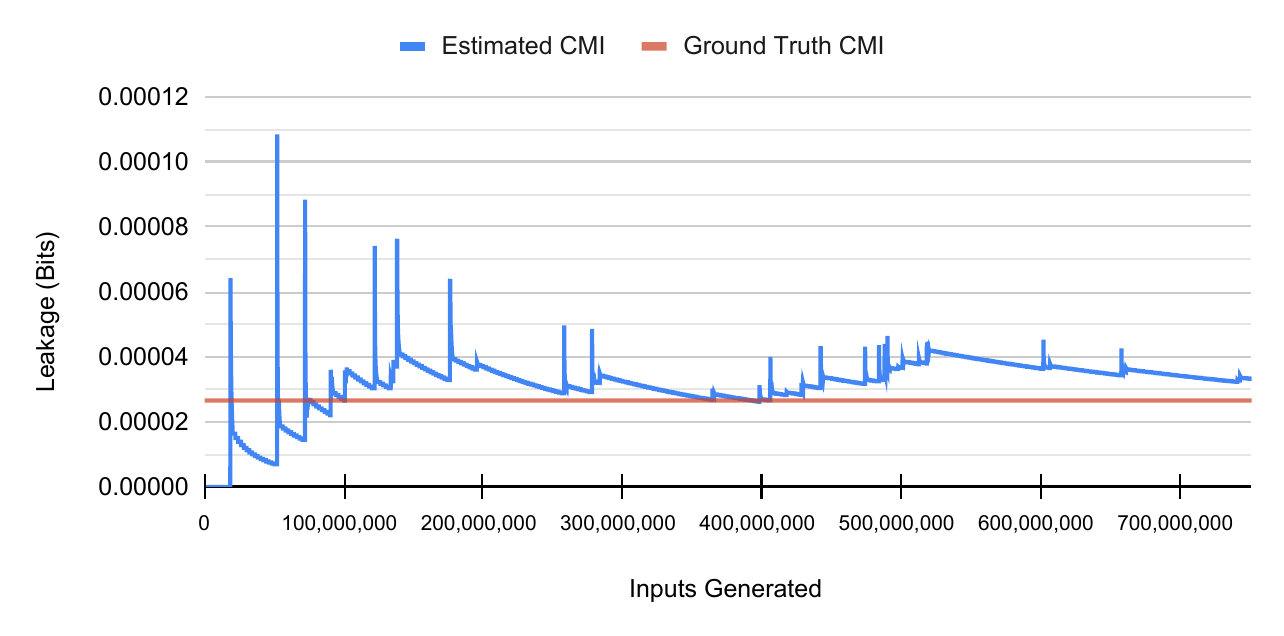}
  
  \caption{Line chart with the blue line showing the estimated CMI value as samples are collected. The red line shows the ground truth value.}
  \label{fig:cmiChart}
\end{figure}

The chart in figure \ref{fig:cmiChart} shows 3 hours worth of fuzzing of the above program (during which time $\approx750$ million inputs were generated and tested).
The steep, near vertical peaks are caused by over estimation of leakage whenever a new violation is found; early on in the campaign (towards the left of the chart) these have a larger effect.

Conditional mutual information could be applied to testing differential privacy applications, or network traffic analysis.
In the case of network traffic analysis, certain publicly readable aspects of say packet size and router destination, act as public input; while the contents (and/or final destination if a VPN is in use) constitute the secret input.

\section{Threats to Validity}

All fuzzing campaigns were run with identical seeds; some were provided by the software developers or original fuzzing harness authors, and where these were not available we used a single simple generic seed that was not specialised to the SUT (i.e. was constructed with no knowledge of the expected input format).
It is possible that the results could differ if the generic starting seed were replaced with different starting seeds.
All internal sources of non-determinism within the fuzzer, such as seeding of the pseudorandom number generators, was kept in place.
Additionally, all of the evaluation has been done with programs that contain known information leak bugs, and we have not tried to discover new, previously unknown ones.
The set of programs on which we did evaluate is limited in size and scope; unfortunately such bugs are hard to find documented, and constructing harnesses for the large multi-user programs that are capable of being affected is an extremely time consuming process.

\section{Conclusion}

In this paper we have introduced \qlfuzz, the first fuzzer capable of quantifying information leaks through program output.

We have derived a new methodology for calculating conditional mutual information in a way that minimises the amount of information that must be stored; we need only to store details for violations.
This measure models an attacker that can observe \emph{public} input (but not influence it) and \emph{public} output.

We have described an approach for detecting whether information leaks come from explicit \emph{secret} inputs, or uninitialised stack or heap memory.
When combined with quantification, this can be useful in identifying the source of the programming error.
We provide an explore and exploit schedule for first finding, and then maximising and quantifying information leaks.
Additionally, we introduce algorithms for detecting direct correspondence between \emph{secret} input and \emph{public} output bits, whereby flipping one results in the flipping of the other.
This trades off accuracy, for reasonable estimation of huge leaks that would be infeasible to quantify through enumeration of outputs.

Finally, we evaluate \qlfuzz, which implements all of the above, on a diverse set of information leaks.
Here we see that it is able to find all of the known information leaks, and provide reasonable lower- and upper-bounds on their quantities -- even for Heartbleed which consists of 278k LoC and leaks over 500k bits.

\bibliographystyle{ACM-Reference-Format}
\bibliography{quantileakfuzzer}


\begin{thebibliography}{33}


\ifx \showCODEN    \undefined \def \showCODEN     #1{\unskip}     \fi
\ifx \showDOI      \undefined \def \showDOI       #1{#1}\fi
\ifx \showISBNx    \undefined \def \showISBNx     #1{\unskip}     \fi
\ifx \showISBNxiii \undefined \def \showISBNxiii  #1{\unskip}     \fi
\ifx \showISSN     \undefined \def \showISSN      #1{\unskip}     \fi
\ifx \showLCCN     \undefined \def \showLCCN      #1{\unskip}     \fi
\ifx \shownote     \undefined \def \shownote      #1{#1}          \fi
\ifx \showarticletitle \undefined \def \showarticletitle #1{#1}   \fi
\ifx \showURL      \undefined \def \showURL       {\relax}        \fi
\providecommand\bibfield[2]{#2}
\providecommand\bibinfo[2]{#2}
\providecommand\natexlab[1]{#1}
\providecommand\showeprint[2][]{arXiv:#2}

\bibitem[Arzt et~al\mbox{.}(2014)]%
        {arzt2014flowdroid}
\bibfield{author}{\bibinfo{person}{Steven Arzt}, \bibinfo{person}{Siegfried
  Rasthofer}, \bibinfo{person}{Christian Fritz}, \bibinfo{person}{Eric Bodden},
  \bibinfo{person}{Alexandre Bartel}, \bibinfo{person}{Jacques Klein},
  \bibinfo{person}{Yves Le~Traon}, \bibinfo{person}{Damien Octeau}, {and}
  \bibinfo{person}{Patrick McDaniel}.} \bibinfo{year}{2014}\natexlab{}.
\newblock \showarticletitle{FlowDroid: precise context, flow, field,
  object-sensitive and lifecycle-aware taint analysis for Android apps}. In
  \bibinfo{booktitle}{\emph{Proceedings of the 35th ACM SIGPLAN Conference on
  Programming Language Design and Implementation}} (Edinburgh, United Kingdom)
  \emph{(\bibinfo{series}{PLDI '14})}. \bibinfo{publisher}{Association for
  Computing Machinery}, \bibinfo{address}{New York, NY, USA},
  \bibinfo{pages}{259–269}.
\newblock
\showISBNx{9781450327848}
\urldef\tempurl%
\url{https://doi.org/10.1145/2594291.2594299}
\showDOI{\tempurl}


\bibitem[Bell et~al\mbox{.}(1976)]%
        {bell1976secure}
\bibfield{author}{\bibinfo{person}{David~E Bell}, \bibinfo{person}{Leonard~J
  La~Padula}, {et~al\mbox{.}}} \bibinfo{year}{1976}\natexlab{}.
\newblock \showarticletitle{Secure computer system: Unified exposition and
  multics interpretation}.
\newblock  (\bibinfo{year}{1976}).
\newblock


\bibitem[Biondi et~al\mbox{.}(2017)]%
        {biondi2017hybrid}
\bibfield{author}{\bibinfo{person}{Fabrizio Biondi}, \bibinfo{person}{Yusuke
  Kawamoto}, \bibinfo{person}{Axel Legay}, {and} \bibinfo{person}{Louis-Marie
  Traonouez}.} \bibinfo{year}{2017}\natexlab{}.
\newblock \showarticletitle{HyLeak: Hybrid Analysis Tool for Information
  Leakage}. In \bibinfo{booktitle}{\emph{Automated Technology for Verification
  and Analysis}}, \bibfield{editor}{\bibinfo{person}{Deepak D'Souza} {and}
  \bibinfo{person}{K.~Narayan~Kumar}} (Eds.). \bibinfo{publisher}{Springer
  International Publishing}, \bibinfo{address}{Cham},
  \bibinfo{pages}{156--163}.
\newblock
\showISBNx{978-3-319-68167-2}


\bibitem[Blackwell et~al\mbox{.}(2025)]%
        {blackwell2025hyperfuzzing}
\bibfield{author}{\bibinfo{person}{Daniel Blackwell}, \bibinfo{person}{Ingolf
  Becker}, {and} \bibinfo{person}{David Clark}.}
  \bibinfo{year}{2025}\natexlab{}.
\newblock \showarticletitle{Hyperfuzzing: black-box security hypertesting with
  a grey-box fuzzer}.
\newblock \bibinfo{journal}{\emph{Empirical Software Engineering}}
  \bibinfo{volume}{30}, \bibinfo{number}{1} (\bibinfo{year}{2025}),
  \bibinfo{pages}{1--28}.
\newblock


\bibitem[B\"{o}hme et~al\mbox{.}(2016)]%
        {AFLFast}
\bibfield{author}{\bibinfo{person}{Marcel B\"{o}hme},
  \bibinfo{person}{Van-Thuan Pham}, {and} \bibinfo{person}{Abhik
  Roychoudhury}.} \bibinfo{year}{2016}\natexlab{}.
\newblock \showarticletitle{Coverage-based Greybox Fuzzing as Markov Chain}. In
  \bibinfo{booktitle}{\emph{Proceedings of the 2016 ACM SIGSAC Conference on
  Computer and Communications Security}} (Vienna, Austria)
  \emph{(\bibinfo{series}{CCS '16})}. \bibinfo{publisher}{Association for
  Computing Machinery}, \bibinfo{address}{New York, NY, USA},
  \bibinfo{pages}{1032–1043}.
\newblock
\showISBNx{9781450341394}
\urldef\tempurl%
\url{https://doi.org/10.1145/2976749.2978428}
\showDOI{\tempurl}


\bibitem[Brennan et~al\mbox{.}(2020)]%
        {brennan2020jvm}
\bibfield{author}{\bibinfo{person}{Tegan Brennan}, \bibinfo{person}{Seemanta
  Saha}, {and} \bibinfo{person}{Tevfik Bultan}.}
  \bibinfo{year}{2020}\natexlab{}.
\newblock \showarticletitle{JVM fuzzing for JIT-induced side-channel
  detection}. In \bibinfo{booktitle}{\emph{Proceedings of the ACM/IEEE 42nd
  International Conference on Software Engineering}} (Seoul, South Korea)
  \emph{(\bibinfo{series}{ICSE '20})}. \bibinfo{publisher}{Association for
  Computing Machinery}, \bibinfo{address}{New York, NY, USA},
  \bibinfo{pages}{1011–1023}.
\newblock
\showISBNx{9781450371216}
\urldef\tempurl%
\url{https://doi.org/10.1145/3377811.3380432}
\showDOI{\tempurl}


\bibitem[Cho et~al\mbox{.}(2020)]%
        {cho2020exploiting}
\bibfield{author}{\bibinfo{person}{Haehyun Cho}, \bibinfo{person}{Jinbum Park},
  \bibinfo{person}{Joonwon Kang}, \bibinfo{person}{Tiffany Bao},
  \bibinfo{person}{Ruoyu Wang}, \bibinfo{person}{Yan Shoshitaishvili},
  \bibinfo{person}{Adam Doup{\'e}}, {and} \bibinfo{person}{Gail-Joon Ahn}.}
  \bibinfo{year}{2020}\natexlab{}.
\newblock \showarticletitle{Exploiting uses of uninitialized stack variables in
  linux kernels to leak kernel pointers}. In \bibinfo{booktitle}{\emph{14th
  USENIX Workshop on Offensive Technologies (WOOT 20)}}.
\newblock


\bibitem[Chothia et~al\mbox{.}(2013)]%
        {chothia2013tool}
\bibfield{author}{\bibinfo{person}{Tom Chothia}, \bibinfo{person}{Yusuke
  Kawamoto}, {and} \bibinfo{person}{Chris Novakovic}.}
  \bibinfo{year}{2013}\natexlab{}.
\newblock \showarticletitle{A tool for estimating information leakage}. In
  \bibinfo{booktitle}{\emph{International Conference on Computer Aided
  Verification}}. Springer, \bibinfo{pages}{690--695}.
\newblock


\bibitem[Clark et~al\mbox{.}(2007)]%
        {clark2007static}
\bibfield{author}{\bibinfo{person}{David Clark}, \bibinfo{person}{Sebastian
  Hunt}, {and} \bibinfo{person}{Pasquale Malacaria}.}
  \bibinfo{year}{2007}\natexlab{}.
\newblock \showarticletitle{A static analysis for quantifying information flow
  in a simple imperative language}.
\newblock \bibinfo{journal}{\emph{Journal of Computer Security}}
  \bibinfo{volume}{15}, \bibinfo{number}{3} (\bibinfo{year}{2007}),
  \bibinfo{pages}{321--371}.
\newblock


\bibitem[Cover and Thomas(2006)]%
        {coverthomas}
\bibfield{author}{\bibinfo{person}{Cover} {and} \bibinfo{person}{Thomas}.}
  \bibinfo{year}{2006}\natexlab{}.
\newblock \bibinfo{booktitle}{\emph{Elements of Information Theory}
  (\bibinfo{edition}{second} ed.)}.
\newblock \bibinfo{publisher}{Wiley}.
\newblock


\bibitem[Denning(1976)]%
        {denning1976lattice}
\bibfield{author}{\bibinfo{person}{Dorothy~E Denning}.}
  \bibinfo{year}{1976}\natexlab{}.
\newblock \showarticletitle{A lattice model of secure information flow}.
\newblock \bibinfo{journal}{\emph{Commun. ACM}} \bibinfo{volume}{19},
  \bibinfo{number}{5} (\bibinfo{year}{1976}), \bibinfo{pages}{236--243}.
\newblock


\bibitem[Fioraldi et~al\mbox{.}(2020)]%
        {fioraldi2020afl++}
\bibfield{author}{\bibinfo{person}{Andrea Fioraldi}, \bibinfo{person}{Dominik
  Maier}, \bibinfo{person}{Heiko Ei{\ss}feldt}, {and} \bibinfo{person}{Marc
  Heuse}.} \bibinfo{year}{2020}\natexlab{}.
\newblock \showarticletitle{$\{$AFL++$\}$: Combining Incremental Steps of
  Fuzzing Research}. In \bibinfo{booktitle}{\emph{14th USENIX Workshop on
  Offensive Technologies (WOOT 20)}}.
\newblock


\bibitem[Fioraldi et~al\mbox{.}(2022)]%
        {LibAFL}
\bibfield{author}{\bibinfo{person}{Andrea Fioraldi},
  \bibinfo{person}{Dominik~Christian Maier}, \bibinfo{person}{Dongjia Zhang},
  {and} \bibinfo{person}{Davide Balzarotti}.} \bibinfo{year}{2022}\natexlab{}.
\newblock \showarticletitle{LibAFL: A Framework to Build Modular and Reusable
  Fuzzers}. In \bibinfo{booktitle}{\emph{Proceedings of the 2022 ACM SIGSAC
  Conference on Computer and Communications Security}} (Los Angeles, CA, USA)
  \emph{(\bibinfo{series}{CCS '22})}. \bibinfo{publisher}{Association for
  Computing Machinery}, \bibinfo{address}{New York, NY, USA},
  \bibinfo{pages}{1051–1065}.
\newblock
\showISBNx{9781450394505}
\urldef\tempurl%
\url{https://doi.org/10.1145/3548606.3560602}
\showDOI{\tempurl}


\bibitem[Godefroid et~al\mbox{.}(2012)]%
        {godefroid2012sage}
\bibfield{author}{\bibinfo{person}{Patrice Godefroid},
  \bibinfo{person}{Michael~Y Levin}, {and} \bibinfo{person}{David Molnar}.}
  \bibinfo{year}{2012}\natexlab{}.
\newblock \showarticletitle{SAGE: whitebox fuzzing for security testing}.
\newblock \bibinfo{journal}{\emph{Commun. ACM}} \bibinfo{volume}{55},
  \bibinfo{number}{3} (\bibinfo{year}{2012}), \bibinfo{pages}{40--44}.
\newblock


\bibitem[Goguen and Meseguer(1982)]%
        {goguen1982security}
\bibfield{author}{\bibinfo{person}{J.~A. Goguen} {and} \bibinfo{person}{J.
  Meseguer}.} \bibinfo{year}{1982}\natexlab{}.
\newblock \showarticletitle{Security Policies and Security Models}. In
  \bibinfo{booktitle}{\emph{1982 IEEE Symposium on Security and Privacy}}.
  \bibinfo{pages}{11--11}.
\newblock
\urldef\tempurl%
\url{https://doi.org/10.1109/SP.1982.10014}
\showDOI{\tempurl}


\bibitem[google({[n.\,d.]})]%
        {fuzzertestsuite}
\bibfield{author}{\bibinfo{person}{google}.}
  \bibinfo{year}{[n.\,d.]}\natexlab{}.
\newblock \bibinfo{title}{fuzzer-test-suite}.
\newblock
\newblock
\newblock
\shownote{\url{https://github.com/google/fuzzer-test-suite}}.


\bibitem[Google(2019)]%
        {aflBasedFuzzers}
\bibfield{author}{\bibinfo{person}{Google}.} \bibinfo{year}{2019}\natexlab{}.
\newblock \bibinfo{title}{afl-based-fuzzers-overview.md}.
\newblock
\newblock
\newblock
\shownote{\url{https://github.com/google/fuzzing/blob/master/docs/afl-based-fuzzers-overview.md}}.


\bibitem[Hamann et~al\mbox{.}(2018)]%
        {hamann2018uniform}
\bibfield{author}{\bibinfo{person}{Tobias Hamann}, \bibinfo{person}{Mihai
  Herda}, \bibinfo{person}{Heiko Mantel}, \bibinfo{person}{Martin Mohr},
  \bibinfo{person}{David Schneider}, {and} \bibinfo{person}{Markus Tasch}.}
  \bibinfo{year}{2018}\natexlab{}.
\newblock \showarticletitle{A uniform information-flow security benchmark suite
  for source code and bytecode}. In \bibinfo{booktitle}{\emph{Nordic Conference
  on Secure IT Systems}}. Springer, \bibinfo{pages}{437--453}.
\newblock


\bibitem[He et~al\mbox{.}(2020)]%
        {he2020ctfuzz}
\bibfield{author}{\bibinfo{person}{Shaobo He}, \bibinfo{person}{Michael Emmi},
  {and} \bibinfo{person}{Gabriela Ciocarlie}.} \bibinfo{year}{2020}\natexlab{}.
\newblock \showarticletitle{ct-fuzz: Fuzzing for Timing Leaks}. In
  \bibinfo{booktitle}{\emph{2020 IEEE 13th International Conference on Software
  Testing, Validation and Verification (ICST)}}. \bibinfo{pages}{466--471}.
\newblock
\urldef\tempurl%
\url{https://doi.org/10.1109/ICST46399.2020.00063}
\showDOI{\tempurl}


\bibitem[Heusser and Malacaria(2010)]%
        {heusser2010quantifying}
\bibfield{author}{\bibinfo{person}{Jonathan Heusser} {and}
  \bibinfo{person}{Pasquale Malacaria}.} \bibinfo{year}{2010}\natexlab{}.
\newblock \showarticletitle{Quantifying information leaks in software}. In
  \bibinfo{booktitle}{\emph{Proceedings of the 26th Annual Computer Security
  Applications Conference}} (Austin, Texas, USA) \emph{(\bibinfo{series}{ACSAC
  '10})}. \bibinfo{publisher}{Association for Computing Machinery},
  \bibinfo{address}{New York, NY, USA}, \bibinfo{pages}{261–269}.
\newblock
\showISBNx{9781450301336}
\urldef\tempurl%
\url{https://doi.org/10.1145/1920261.1920300}
\showDOI{\tempurl}


\bibitem[Hocevar(2007)]%
        {hocevar_2007}
\bibfield{author}{\bibinfo{person}{Sam Hocevar}.}
  \bibinfo{year}{2007}\natexlab{}.
\newblock \bibinfo{title}{zzuf - multi-purpose fuzzer}.
\newblock
\newblock
\newblock
\shownote{\url{http://caca.zoy.org/wiki/zzuf}}.


\bibitem[Kim et~al\mbox{.}(2014)]%
        {kim2014flipping}
\bibfield{author}{\bibinfo{person}{Yoongu Kim}, \bibinfo{person}{Ross Daly},
  \bibinfo{person}{Jeremie Kim}, \bibinfo{person}{Chris Fallin},
  \bibinfo{person}{Ji~Hye Lee}, \bibinfo{person}{Donghyuk Lee},
  \bibinfo{person}{Chris Wilkerson}, \bibinfo{person}{Konrad Lai}, {and}
  \bibinfo{person}{Onur Mutlu}.} \bibinfo{year}{2014}\natexlab{}.
\newblock \showarticletitle{Flipping bits in memory without accessing them: An
  experimental study of DRAM disturbance errors}.
\newblock \bibinfo{journal}{\emph{ACM SIGARCH Computer Architecture News}}
  \bibinfo{volume}{42}, \bibinfo{number}{3} (\bibinfo{year}{2014}),
  \bibinfo{pages}{361--372}.
\newblock


\bibitem[Kocher et~al\mbox{.}(2019)]%
        {kocher2019spectre}
\bibfield{author}{\bibinfo{person}{Paul Kocher}, \bibinfo{person}{Jann Horn},
  \bibinfo{person}{Anders Fogh}, \bibinfo{person}{Daniel Genkin},
  \bibinfo{person}{Daniel Gruss}, \bibinfo{person}{Werner Haas},
  \bibinfo{person}{Mike Hamburg}, \bibinfo{person}{Moritz Lipp},
  \bibinfo{person}{Stefan Mangard}, \bibinfo{person}{Thomas Prescher},
  {et~al\mbox{.}}} \bibinfo{year}{2019}\natexlab{}.
\newblock \showarticletitle{Spectre attacks: Exploiting speculative execution}.
  In \bibinfo{booktitle}{\emph{2019 IEEE Symposium on Security and Privacy
  (SP)}}. IEEE, \bibinfo{pages}{1--19}.
\newblock


\bibitem[Lipp et~al\mbox{.}(2018)]%
        {lipp2018meltdown}
\bibfield{author}{\bibinfo{person}{Moritz Lipp}, \bibinfo{person}{Michael
  Schwarz}, \bibinfo{person}{Daniel Gruss}, \bibinfo{person}{Thomas Prescher},
  \bibinfo{person}{Werner Haas}, \bibinfo{person}{Anders Fogh},
  \bibinfo{person}{Jann Horn}, \bibinfo{person}{Stefan Mangard},
  \bibinfo{person}{Paul Kocher}, \bibinfo{person}{Daniel Genkin},
  {et~al\mbox{.}}} \bibinfo{year}{2018}\natexlab{}.
\newblock \showarticletitle{Meltdown: Reading kernel memory from user space}.
  In \bibinfo{booktitle}{\emph{27th USENIX Security Symposium (USENIX Security
  18)}}. \bibinfo{pages}{973--990}.
\newblock


\bibitem[Mathis et~al\mbox{.}(2017)]%
        {bjorn2017detecting}
\bibfield{author}{\bibinfo{person}{Björn Mathis}, \bibinfo{person}{Vitalii
  Avdiienko}, \bibinfo{person}{Ezekiel~O. Soremekun}, \bibinfo{person}{Marcel
  Böhme}, {and} \bibinfo{person}{Andreas Zeller}.}
  \bibinfo{year}{2017}\natexlab{}.
\newblock \showarticletitle{Detecting information flow by mutating input data}.
  In \bibinfo{booktitle}{\emph{2017 32nd IEEE/ACM International Conference on
  Automated Software Engineering (ASE)}}. \bibinfo{pages}{263--273}.
\newblock
\urldef\tempurl%
\url{https://doi.org/10.1109/ASE.2017.8115639}
\showDOI{\tempurl}


\bibitem[Meng and Smith(2011)]%
        {meng2011calculating}
\bibfield{author}{\bibinfo{person}{Ziyuan Meng} {and} \bibinfo{person}{Geoffrey
  Smith}.} \bibinfo{year}{2011}\natexlab{}.
\newblock \showarticletitle{Calculating bounds on information leakage using
  two-bit patterns}. In \bibinfo{booktitle}{\emph{Proceedings of the ACM
  SIGPLAN 6th Workshop on Programming Languages and Analysis for Security}}
  (San Jose, California) \emph{(\bibinfo{series}{PLAS '11})}.
  \bibinfo{publisher}{Association for Computing Machinery},
  \bibinfo{address}{New York, NY, USA}, Article \bibinfo{articleno}{1},
  \bibinfo{numpages}{12}~pages.
\newblock
\showISBNx{9781450308304}
\urldef\tempurl%
\url{https://doi.org/10.1145/2166956.2166957}
\showDOI{\tempurl}


\bibitem[Mesecan et~al\mbox{.}(2022)]%
        {MesecanBCCP22}
\bibfield{author}{\bibinfo{person}{Ibrahim Mesecan}, \bibinfo{person}{Daniel
  Blackwell}, \bibinfo{person}{David Clark}, \bibinfo{person}{Myra~B. Cohen},
  {and} \bibinfo{person}{Justyna Petke}.} \bibinfo{year}{2022}\natexlab{}.
\newblock \showarticletitle{Keeping Secrets: Multi-objective Genetic
  Improvement for Detecting and Reducing Information Leakage}. In
  \bibinfo{booktitle}{\emph{37th {IEEE/ACM} International Conference on
  Automated Software Engineering, {ASE} 2022, Rochester, MI, USA, October
  10-14, 2022}}. \bibinfo{publisher}{{ACM}}, \bibinfo{pages}{61:1--61:12}.
\newblock


\bibitem[Nilizadeh et~al\mbox{.}(2019)]%
        {nilizadeh2019diffuzz}
\bibfield{author}{\bibinfo{person}{Shirin Nilizadeh}, \bibinfo{person}{Yannic
  Noller}, {and} \bibinfo{person}{Corina~S. Pasareanu}.}
  \bibinfo{year}{2019}\natexlab{}.
\newblock \showarticletitle{DifFuzz: Differential Fuzzing for Side-Channel
  Analysis}. In \bibinfo{booktitle}{\emph{2019 IEEE/ACM 41st International
  Conference on Software Engineering (ICSE)}}. \bibinfo{pages}{176--187}.
\newblock
\urldef\tempurl%
\url{https://doi.org/10.1109/ICSE.2019.00034}
\showDOI{\tempurl}


\bibitem[Noller and Tizpaz-Niari(2021)]%
        {noller2021qfuzz}
\bibfield{author}{\bibinfo{person}{Yannic Noller} {and} \bibinfo{person}{Saeid
  Tizpaz-Niari}.} \bibinfo{year}{2021}\natexlab{}.
\newblock \showarticletitle{QFuzz: quantitative fuzzing for side channels}. In
  \bibinfo{booktitle}{\emph{Proceedings of the 30th ACM SIGSOFT International
  Symposium on Software Testing and Analysis}} (Virtual, Denmark)
  \emph{(\bibinfo{series}{ISSTA 2021})}. \bibinfo{publisher}{Association for
  Computing Machinery}, \bibinfo{address}{New York, NY, USA},
  \bibinfo{pages}{257–269}.
\newblock
\showISBNx{9781450384599}
\urldef\tempurl%
\url{https://doi.org/10.1145/3460319.3464817}
\showDOI{\tempurl}


\bibitem[Pasqua et~al\mbox{.}(2024)]%
        {pasqua2024hypertesting}
\bibfield{author}{\bibinfo{person}{Michele Pasqua}, \bibinfo{person}{Mariano
  Ceccato}, {and} \bibinfo{person}{Paolo Tonella}.}
  \bibinfo{year}{2024}\natexlab{}.
\newblock \showarticletitle{Hypertesting of Programs: Theoretical Foundation
  and Automated Test Generation}. In \bibinfo{booktitle}{\emph{Proceedings of
  the IEEE/ACM 46th International Conference on Software Engineering}}.
  \bibinfo{pages}{1--12}.
\newblock


\bibitem[Phan and Malacaria(2014)]%
        {phan2014abstract}
\bibfield{author}{\bibinfo{person}{Quoc-Sang Phan} {and}
  \bibinfo{person}{Pasquale Malacaria}.} \bibinfo{year}{2014}\natexlab{}.
\newblock \showarticletitle{Abstract model counting: a novel approach for
  quantification of information leaks}. In
  \bibinfo{booktitle}{\emph{Proceedings of the 9th ACM Symposium on
  Information, Computer and Communications Security}} (Kyoto, Japan)
  \emph{(\bibinfo{series}{ASIA CCS '14})}. \bibinfo{publisher}{Association for
  Computing Machinery}, \bibinfo{address}{New York, NY, USA},
  \bibinfo{pages}{283–292}.
\newblock
\showISBNx{9781450328005}
\urldef\tempurl%
\url{https://doi.org/10.1145/2590296.2590328}
\showDOI{\tempurl}


\bibitem[Zalewski(2014)]%
        {zalewski}
\bibfield{author}{\bibinfo{person}{Micha{\l} Zalewski}.}
  \bibinfo{year}{2014}\natexlab{}.
\newblock \bibinfo{title}{american fuzzy lop (2.52b)}.
\newblock
\newblock
\newblock
\shownote{\url{http://lcamtuf.coredump.cx/afl/}}.


\bibitem[Zdancewic(2004)]%
        {SZ04}
\bibfield{author}{\bibinfo{person}{Steve Zdancewic}.}
  \bibinfo{year}{2004}\natexlab{}.
\newblock \showarticletitle{Challenges for information-flow security}. In
  \bibinfo{booktitle}{\emph{Proceedings of the 1st International Workshop on
  Programming Language Interference and Dependence (PLID’04)}}.
\newblock


\end{thebibliography}

\end{document}